\documentclass[journal=jctcce,manuscript=article]{achemso} 
\usepackage[T1]{fontenc} 
\usepackage[version=3]{mhchem} 
\usepackage[dvipsnames]{xcolor}
\usepackage{graphicx,subcaption}
\usepackage{amssymb}
\usepackage{overpic}
\usepackage{soul}
\usepackage{bm}
\usepackage{hyperref}
\usepackage{graphicx}

\usepackage{braket}

\newif\ifhighlightchanges
\highlightchangesfalse

\newcommand{\mycomment}[1]{{\color{red} #1 }}

\definecolor{darkblue}{HTML}{003D6D}

\newcommand{\bR}{{\bm R}}
\newcommand{\br}{{\bm r}}
\newcommand{\bg}{{\bm g}}
\newcommand{\bP}{{\bm P}}

\newcommand{\hC}{{\hat{C}}}

\newcommand{\mW}{{\mathcal{W}}}

\newcommand{\bp}{{\bm p}}

\newcommand{\bd}{{\bm d}}
\newcommand{\bzeta}{{\bm \zeta}}

\newcommand{\bV}{{\bm V}}

\newcommand{\hH}{\hat H}
\newcommand{\hT}{\hat T}

\newcommand{\hY}{\hat Y}
\newcommand{\hA}{\hat A}
\newcommand{\hB}{\hat B}
\newcommand{\hJ}{\hat J}
\newcommand{\hU}{\hat U}

\newcommand{\hO}{\hat O}
\newcommand{\hV}{\hat V}

\newcommand{\hLambda}{\hat \Lambda}
\newcommand{\hP}{\hat P}

\newcommand{\hL}{\hat L}
\newcommand{\hK}{\hat K}
\newcommand{\hbp}{{\hat{\bm p}}}
\newcommand{\hbV}{{\hat{\bm V}}}
\newcommand{\hbP}{{\hat{\bm P}}}
\newcommand{\hbH}{{\hat{\bm H}}}
\newcommand{\hbd}{{\hat{\bm d}}}
\newcommand{\hbr}{{\hat{\bm r}}}

\newcommand{\hbR}{{\hat{\bm R}}}

\newcommand{\bH}{{\bm H}}

\newcommand{\hGamma}{{\hat{\Gamma}}}
\newcommand{\hbGamma}{{\hat{{\bm \Gamma}}}}

\usepackage{caption}

\title{Recovering Exact Vibrational Energies Within a Phase Space Electronic Structure Framework}

\author{Xinchun Wu}
\affiliation{Department of Chemistry, Princeton University, Princeton, New Jersey 08544, USA}
\alsoaffiliation{Department of Chemistry, University of Pennsylvania, Philadelphia, Pennsylvania 19104, USA}
\author{Xuezhi Bian}
\affiliation{Department of Chemistry, Princeton University, Princeton, New Jersey 08544, USA}
\author{Jonathan Rawlinson}
\affiliation{School of Science \& Technology, Nottingham Trent University, Nottingham
NG1 4FQ, UK}
\author{Robert G. Littlejohn}
\affiliation{Department of Physics, University of California, Berkeley, California 94720, USA}
\author{Joseph E. Subotnik}
\email{subotnik@princeton.edu}
\affiliation{Department of Chemistry, Princeton University, Princeton, New Jersey 08544, USA}
\alsoaffiliation{Department of Chemistry, University of Pennsylvania, Philadelphia, Pennsylvania 19104, USA}
\date{May 2025}

\begin{document}

\begin{abstract}
  In recent years, there has been a push to go beyond Born-Oppenheimer theory and build electronic states from a phase space perspective, i.e. parameterize electronic states by both nuclear position ($\bR$) and nuclear momentum $(\bP)$.  Previous empirical studies have demonstrated that such approaches can yield improved single-surface observables, including vibrational energies, electronic momenta, and vibrational circular dichroism spectra. 
 That being said, unlike the case of BO theory, there is no unique phase space electronic Hamiltonian, nor any theory for using phase space eigenvectors (as opposed to BO eigenvectors) so as to recover exact quantum vibrational eigenvalues. As such, one might consider such phase space approaches {\em ad hoc}. To that end, here we show how to formally extract exact quantum energies from a coupled nuclear-electronic Hamiltonian using perturbation theory on top of a phase space  
 electronic framework. 
Thus, while we cannot isolate an ``optimal'' phase space electronic Hamiltonian, this work does justify a  phase space electronic structure approach by offering a rigorous framework for correcting the zeroth order phase space electronic states.
\end{abstract}

\maketitle

\section{Introduction: Vibrational Energies as a Probe of Potential Energy Surfaces}

Identifying vibrational energies is one of the chemist's most useful tools, which aids in identifying functional groups\cite{colthup2012introduction}, molecular structure and local temperature\cite{frontiera:2016:jpcl:raman_hot}.  Moreover, the theory of how to calculate vibrational modes is effectively as old as quantum mechanics itself.\cite{heitler1927wechselwirkung}
Consider a Hamiltonian for a molecular or material system composed of nuclei and electrons:
\begin{eqnarray}
\label{eqn:Htot}
    \hH & =& \hT_n + \hH_{el} \\
     \hT_n &=& \sum_{I=1}^{N_n} \frac{\hbP_I^2}{2M_I} \\
     \hH_{el} &=& \sum_{i=1}^{N_e} \frac{\hbp_i^2}{2m_e} 
    -    \sum_{I=1}^{N_n} \sum_{i = 1}^{N_e} \frac{Q_Ie} {4\pi \epsilon_0 \left| \hbr_i - \hbR_I \right|} \nonumber \\
    & & \; \; \; \; \; \; \; \; 
    +    \sum_{I=1}^{N_n} \sum_{J = I+1}^{N_n} \frac{Q_I Q_J}{4\pi \epsilon_0 \left| \hbR_J - \hbR_I \right|} +
    \sum_{i=1}^{N_e} \sum_{j = i+1}^{N_e} \frac{e^2}{4\pi \epsilon_0 \left| \hbr_i - \hbr_j \right|} 
\end{eqnarray}
Here, we let $\bp_i$ be the momentum of electron $i$, while $Q_I$ is the charge of nucleus $I$, $\bR_I$ is the position of nucleus $I$, and $M_I$ is the mass of nucleus $I$.
The electron mass is $m_e$
Due to the extreme computational cost in diagonalizing $\hH_{tot}$, it is standard to investigate chemical systems through the {\em Born-Oppenheimer framework}\cite{born1996dynamical, cederbaum:review:conicalbook}. Within this framework,   one begins  any analysis by diagonalizing the  electronic Hamiltonian ($\hH_{el}$) to generate a basis of electronic adiabatic states:
 \begin{eqnarray}
 \label{eqn:Hel}
\hat{H}_{el} \Phi_j(\br;\bR)  = V_j(\bR) \Phi_j(\br;\bR)
 \end{eqnarray}
 Here, $V_0(\bR)$ denotes the ground state {\em potential energy surface} for the nuclei. 
 Thereafter, in the basis of adiabatic electronic states, the full Hamiltonian takes the form:
\begin{eqnarray}
\hH = \sum_{njkA} \ket{n} \frac{(\hbP_A \delta_{nk} - i \hbar \hbd_{nk}^A) \cdot (\hbP_A \delta_{kj} - i \hbar \hbd_{kj}^A)} {2M}  \bra{j}  + \sum_{nj}\hbV_{nj}\ket{n}\bra{j}
\label{eq:stdbo:1}
\end{eqnarray}
 where the matrix of derivative couplings are defined as:
 \begin{eqnarray}
\label{eq:defdij}
    \bd_{jk}^A = \bra{\Phi_j} \nabla_{{R}_A} \ket{\Phi_k}
\end{eqnarray}

Next, 
 according to the BO {\em approximation}, one ignores the derivative couplings and generates the stationary states of the system by solving the nuclear Schrodinger equation on a single surface. In particular, for the ground state  (0) energy, one diagonalizes the operator:
 \begin{eqnarray}
  \label{eqn:HBO}
\hat{H}_{\mathrm{BO}}=\frac{\hat{\bP}^2}{2 M}+V_0(\hat{\bR})
 \end{eqnarray}
 
One of the most important features of the BO framework is the capacity to calculate and characterize quantum states according to the nature of the excitation, e.g. electronic, vibrational, and rotational excitations. In particular, for many chemical systems, one can diagonalize $\hat{H}_{\mathrm{BO}}$ (rather than $\hat{H}$) so as to generate vibrational excitations that agree well with experiment\cite{bernath2020spectra}.
 Thus, the agreement between BO vibrational energies and experimental spectra has always been a resounding endorsement of the notion of potential energy surfaces.
That being said, however, it must be noted that within the realm of high-resolution spectroscopy, it has long been appreciated that one can measure deviations from the BO approximation\cite{lefebvre2004spectra}.  Indeed, a host of corrections must be introduced if one seeks to match BO vibrational energies with high-resolution spectroscopy.\cite{bunker1980effect,bunker2006molecular,schwenke2001beyond, scherrer2017mass} In other words,  the BO approximation is (like all approximations) approximate;
it is impossible to absolutely separate electronic and vibrational excitations.  Indeed, this statement is not controversial for anyone familiar with the curl conditions for diabatic states.\cite{mbaer:1975:cpl, meadtruhlar:1982:conditions_diabatic}

At this point, one must wonder: 
if one seeks theory to match high-resolution spectroscopy, is there a meaningful and efficient alternative BO framework? More bluntly, 
if one seeks to characterize vibrational energies within a single surface approximation, 
are we confident that the BO surfaces are the optimal surfaces? Is there perhaps a richer and more complete potential energy surface? Even more broadly, are there other possible basis sets and frameworks for expanding coupled nuclear-electronic wavefunctions \cite{bubin2013born}?

To that end, over the last few years, our research group has vigorously investigated the possibility of replacing the BO framework and the BO electronic Hamiltonian ($\hH_{el}(\bR)$) with a phase space framework and a phase space (PS) electronic Hamiltonian $\hH_{PS}(\bR,\bP)$ which is parameterized by both nuclear $\bR$ and $\bP$.  To be specific, we have posited a phase space Hamiltonian of the form
\begin{eqnarray}
\label{eqn:Hps}
    \hH_{PS}  \equiv  
    \sum_I \frac{\left(\bP_I - i \hbar \hbGamma_I\right)^2}{2M_I} + \hH_{el}
\end{eqnarray}
where $\hbGamma$ is a one-electron operator that satisfies  key translation and rotational conditions that are obeyed by the adiabatic derivative coupling.  For the exact form of $\hbGamma$, please see Refs. \citenum{tao2025basis, bian:2025:jctc:wigner_vibrations}, as well as Eq. \ref{eq:HPS:model} below. (Note, though, that all formal work below will not depend on the particular form of  $\hbGamma$.) 

In all cases to date, a phase space electronic Hamiltonian of the form in Eq. \ref{eqn:Hps} has outperformed the BO electronic Hamiltonian, often with very meaningful implications. In particular, a PS approach outperforms BO as far as:

\begin{enumerate}
\item \label{itemp} PS theory offers meaningful (nonzero) electronic momenta for dynamics along the ground state\cite{tao2025basis,tao2024practical}, whereas BO theory predicts zero electronic momenta, a fact which has motivated the theory for quite some time\cite{nafie1983adiabatic, patchkovskii:2012:jcp:electronic_current, takatsuka:2021:jcp:flux_conservation}.

\item PS conserves the total linear and angular momentum\cite{tao2025basis,tian:2024:jcp:erf}. Formally BO theory also conserves the total nuclear momentum\cite{littlejohn:2023:jcp:angmom} and classical BO dynamics should carry a Berry force\cite{bian2023total}, but for systems with an even number of electrons and a non-degenerate time-reversal ground state, the Berry force vanishes and BO sets the electronic momentum to zero (incorrectly). As such, within BO theory, in certain cases, the total momentum is no different from the nuclear momentum.

\item PS theory easily recovers an accurate VCD signal with only a single perturbation \cite{duston2024phase,tao2024electronic}, whereas one must perform magnetic field perturbation theory with a double perturbation within BO theory\cite{stephens:1985:jpcc_vcd}.

\item \label{itemvib} PS offers improved vibrational energies. Often times, these improvements are small, but in cases of artificially large electronic masses (which break the BO approximation), the improvements can be substantial. \cite{bian:2025:jctc:wigner_vibrations}

\item Polkovnikov {\em et al} have shown that PS approaches have analogies in the classical world, offering a new twist on how to derive smooth (rather than noisy) slow variables\cite{d2014emergent, polkovnikov:2025:moving_boa}.
\end{enumerate}

For the purposes of this paper let us now discuss item \ref{itemvib} in more detail.  Starting from a potential energy surface that depends on both $\bP$ and $\bR$, $V(\bR,\bP)$, there is no obvious  path to generate a nuclear quantum Hamiltonian.  After all, $\bR$ and $\bP$ do not commute and thus the $\bR$ and $\bP$ in $V(\bR,\bP)$ cannot be regarded as  quantum operators.  That being said, in Ref. \citenum{bian:2025:jctc:wigner_vibrations}, in order to extract vibrational energies, we performed a Weyl transform,
\begin{equation}
\label{eqn:weyl}
\langle \bR| \hat{H}_{P S}\left|\bR^{\prime}\right\rangle=\int d \bP e^{\frac{i}{\hbar} \bP \cdot\left(\bR-\bR^{\prime}\right)} E_{P S}\left(\frac{\bR+\bR^{\prime}}{2}, \bP\right)
\end{equation}
and then diagonalized the resulting $\langle \bR| \hat{H}_{P S}\left|\bR^{\prime}\right\rangle$.

Despite the successes described above (and especially diagonalizing Eq. \ref{eqn:weyl} for vibrational energies \cite{bian:2025:jctc:wigner_vibrations}), the introduction  of a phase space electronic Hamiltonian inevitably raises two criticisms. First, unlike $\hH_{el}$ in Eq. \ref{eqn:Hel}, there is no unique phase space electronic Hamiltonian because $\hbGamma$ in Eq. \ref{eqn:Hps} is not unique.  Second, even though diagonalization of Eq. \ref{eqn:Hps} on a single surface yields better performing vibrational energies than BO theory does, one could argue that diagonalizing Eq. \ref{eqn:Hps} is {\em ad hoc} because the result is not systematically improvable. In other words, no one has yet shown how to generate an exact coupled nuclear-electronic eigenfunction of Eq. \ref{eqn:Htot} starting from phase-space electronic states (whereas one can work with BO electronic states for an exact diagonalization).

In what follows below, we will now address the second criticism above. Even without a unique phase space electronic Hamiltonian, we will show below that one {\em can} use a phase space electronic Hamiltonian approach as a starting point for an exact diagonalization of $H_{tot}$. Our approach will follow ideas set forth by Blount\cite{blount1962} originally, and later formalized by Littlejohn  and Flynn\cite{littlejohn1991,littlejohn:flynn:1991:pra:coriolis, littlejohn:2024:jcp:moyal} and Teufel\cite{teufel2003, matyus2019}.  The basic idea is to  use perturbation theory in tandem within a Wigner representation to generate exact eigenstates with the desired properties.
Within the theoretical chemistry community, Wigner-Weyl transforms are very well known in the dynamics community and can be used to develop nonadiabatic dynamics techniques\cite{martens:1997:partwig,martens:1998:partwig,kapral:1999:jcp, kapral:2009:burghardt,kelly:2010:jcp,kapral:2000:jcp, coker:2012:iterative,geva:shi:2003:semiclassical_relaxation:jpca}; here, however, we will show that the same transform can be incredibly helpful for designing new techniques in electronic structure. 
Thus,  in the end, the present paper  justifies using a phase space electronic Hamiltonian, putting the latter on a very rigorous (and even practical) footing. 

An outline of this paper is as follows. In Sec.\ref{sec:boreview} of this paper, we will review BO theory, showing first how the standard expansion can be derived within Wigner theory and then  deriving the Littlejohn-Flynn perturbative expansion.  In Sec.\ref{sec:ps}, we will then review phase space electronic structure theory and formally derive a perturbative expansion that allows one to recover an exact eigenstate starting within a phase space framework. In Sec.\ref{sec:model}, we demonstrate that the present formalism indeed works numerically using a  model Hamiltonian. In Sec.\ref{sec:mpandha}, we discuss two relevant nuances -- the mass polarization term and the harmonic limit.  We conclude in Sec.\ref{sec:conclusion}.

Finally, a word about notation. In this manuscript, it will be crucial to distinguish between operators and scalars, as applied to both electronic and nuclear degrees of freedom. In our introduction above, we have used reasonably standard notation whereby all operators (nuclear or electronic) are written with hats. That being said,
given that we will work with Wigner transforms below and nuclei will be treated differently from electrons, the manuscript is written for a system with one nuclear degree of freedom (though extensions are obvious). Moreover, henceforward, we will change notation and use hats ($\hat{U}$) exclusively for  electronic operators, and  boldface (${\bf U}$) exclusively for nuclear operators.  While this notation is not standard, we believe it is the simplest means to communicate the final answer and intuition. Thus, going forward, our single nuclear degrees of freedom has classical phase space coordinates $(R,P)$ and the corresponding quantum operators are $(\bR,\bP)$.
Lastly, a word about superscripts and perturbation theory.
If we perform perturbation theory, one typically writes the final energy as:
\begin{eqnarray}
    E = E^{(0)} + E^{(1)} + E^{(2)} + \ldots
\end{eqnarray}
For our purposes, it will also be helpful to have a shorthand for the intermediate sums. Thus, we will use the tilde notation $\sim$  to signify such intermediate sums:
\begin{eqnarray}
    \tilde{E}^{(0)} &=& E^{(0)}  \\
   \tilde{E}^{(1)} &=& E^{(0)}  + E^{(1)} \\
      \tilde{E}^{(2)} &=& E^{(0)}  + E^{(1)} + E^{(2)} 
\end{eqnarray}
and so forth.

\section{Vibrational Energies through the BO Framework}
\label{sec:boreview}
For a chemical problem with one nuclear degree of freedom, the adiabatic representation from Eq. \ref{eq:stdbo:1} above has the following matrix elements:
\begin{eqnarray}
\bH_{nj} = \sum_{k}  (\bP \delta_{nk} - i \hbar \bd_{nk})\frac{1}{2M} (\bP \delta_{kj} - i \hbar \bd_{kj})   + \bV_{nj}
\label{eq:stdbo:2}
\end{eqnarray}
Note that the sum over states in  Eqs. \ref{eq:stdbo:1} and  \ref{eq:stdbo:2}  must extend over a complete set of states $n,j,k$ for an exact representation of the Hamiltonian.

Now, note that we can consider Eq. \ref{eq:stdbo:1} formally as an expansion in $\hbar$:
\begin{eqnarray}
\label{eq:BO:expand_in_hbar}
\hbH = \hbH_{BO}^{(0)} + \hbH_{BO}^{(1)} + \hbH_{BO}^{(2)}
\end{eqnarray}
with corresponding matrix elements:
\begin{eqnarray}
\label{eqn:Hbo^0}
(\hbH_{BO}^{(0)})_{nj} &=&
 \delta_{nj} \frac{\bP^2}{2M}  + \bV_{nj}\\
\label{eqn:Hbo^1}
(\hbH_{BO}^{(1)})_{nj} &=&  \frac{- i \hbar   \bP \cdot \bd_{nj} - i \hbar  \bd_{nj} \cdot \bP}{2M} 
\\
\label{eqn:Hbo^2}
(\hbH_{BO}^{(2)})_{nj} &=&  \sum_{k} \frac{-\hbar^2 \bd_{nk} \cdot \bd_{kj}}{2M}  
\equiv -\hbar^2 \frac{\bzeta_{nj}}{2M}
\end{eqnarray}

Note that one can develop a slightly different 
 $\hbar$ expansion
by writing the matrix elements in Eq.  \ref{eq:stdbo:2}   as:
\begin{eqnarray}
\label{eqn:bodifferent}
    \bH_{nj} =\delta_{nj}\frac{\bP^2}{2M}-i \hbar \bd_{nj}\frac{\bP}{M} - \hbar^2 \frac{\bg_{nj}}{2M} + \bV_{nj}
\end{eqnarray}
where $\bg$ is the second derivative coupling,
\begin{eqnarray}
    \bg_{nj} = \left<n \middle|  \nabla_{R}^2 j \right> 
\end{eqnarray}
Eq. \ref{eq:stdbo:2} and Eq. \ref{eqn:bodifferent} are identical as can be shown by using the identities (with $f$ be an arbitrary trial function)
\begin{eqnarray}
    \left< n \middle |  \nabla_{R}^2 j \right>  &= & \nabla_{R} \left< n \middle | \nabla_{R} j \right> -\braket{\nabla_{R} n | \nabla_{R} j} \\
 \bP \bd_{nj} f & = & f (\bP  \bd_{nj}) + \bd_{nj} (\bP f) = -i \hbar f(\nabla_{R} \left< n \middle| \nabla_R  j \right>)  -i \hbar \bd_{nj} ( \nabla_{R} f)
\end{eqnarray}
and plugging  into Eq. \ref{eq:stdbo:2}:
\begin{eqnarray}
   \bH_{nj} &= & \delta_{nj}\frac{\bP^2}{2M} + \frac{- i \hbar  \bP \cdot \bd_{nj} - i \hbar  \bd_{nj} \cdot \bP}{2M} -\sum_k \frac{\hbar^2 \bd_{nk} \bd_{kj}}{2M} + \bV_{nj}\\
  & = & \delta_{nj}\frac{\bP^2}{2M} +\frac{- i \hbar \left(-i \hbar (\nabla \bra{n}\nabla \ket{j}) + \bd_{nj} \cdot \bP \right) - i \hbar  \bd_{nj} \cdot \bP -
  \sum_k \hbar^2 \bd_{nk} \bd_{kj} }{2M} + \bV_{nj} \\
   & = & \delta_{nj}\frac{\bP^2}{2M} +\frac{-i \hbar \bd_{nj} \bP }{M} +\frac{-\hbar^2 (\braket{\nabla n | \nabla j} + \bra{n} \nabla_{\bR}^2 \ket{j}) - 
   \sum_k \hbar^2 \bd_{nk} \bd_{kj}}{2M} + \bV_{nj}\\
   & = & \delta_{nj}\frac{\bP^2}{2M} +\frac{-i \hbar \bd_{nj} \bP }{M} -\hbar^2 \frac{ \sum_k (\braket{\nabla n | k}\braket{k | \nabla j} + \bra{n} \nabla_{\bR}^2 \ket{j}) +  \sum_k \braket{n | \nabla k} \braket{k | \nabla j} }{2M} + \bV_{nj}
   \nonumber 
   \\ 
   \\
   & = & \delta_{nj}\frac{\bP^2}{2M} -\frac{i \hbar \bd_{nj} \bP }{M} - \hbar^2 \frac{\bg_{nj}}{2M} + \bV_{nj}
\end{eqnarray}
For our purposes, 
Eq. \ref{eq:stdbo:2} is preferable to Eq. \ref{eqn:bodifferent} because, with the former, each of the three terms in the Hamiltonian is hermitian. In any event, 
the BO approximation arises by neglecting the linear and quadratic terms in $\hbar$:

\begin{eqnarray}
 \hbH_{BO} =  \hbH_{BO}^{(0)} 
\end{eqnarray}

\subsection{A Wigner Derivation of the BO Framework}
\label{sec:wignerbo}
While not commonly used, Eq. \ref{eq:stdbo:1} can also be derived through a Wigner transform. Recall the definitions of Wigner ($\mathcal{W}$) and Weyl  ($\mathcal{W}^{-1}$) transformations:
\begin{eqnarray}
    \hO_W &=& \mathcal{W}(\hO) \\
    \hO_W(R,P) &=& \int d\bm R' \bra{\bm R + \frac {\bm R'} 2 } \hat O \ket{\bm R - \frac {\bm R'} 2 } e^{-\frac i \hbar \bm R' \cdot \bm P} \\
    \hO &=& \mathcal{W}^{-1}(\hO_W) \\
    \label{eqn:weyltransform}
     \left< R \middle| \hO \middle| R' \right> &=& \int \frac {d\bm P} {2\pi\hbar} e^{\frac i\hbar \bm P \cdot (\bm R - \bm R')} \hat O_W\left(\frac {\bm R + \bm R'} 2, \bm P\right) 
\end{eqnarray}
Recall also the fact that, for two operators, $\hO_1$ and $\hO_2$, 
\begin{eqnarray}
\label{eqn:starp}
    \left( \hO_1 \hO_2 \right)_W =      (\hO_1)_W * (\hO_2)_W 
\end{eqnarray}
where the star (Moyal) product is:
\begin{eqnarray}
* = \exp\left(\frac{i\hbar}{2} \left(
\overleftarrow{\nabla}_R
\overrightarrow{\nabla}_P - 
\overleftarrow{\nabla}_P
\overrightarrow{\nabla}_R 
\right) \right)
\end{eqnarray}

Consider now the partial nuclear Wigner transform of the total Hamiltonian in Eq. \ref{eqn:Htot}:
\begin{eqnarray}
    \hH_W = \frac{P_W^2}{2M} + \hV_W
\end{eqnarray}
The usual BO approximation comes from the electronic diagonalization of $\hH_W$
\begin{eqnarray}
        \hH_W = \hU_W \hat{\Lambda}^{BO}_W \hU^{\dagger}_W 
\end{eqnarray}
and then taking the inverse Weyl transformation of $\hat{\Lambda}_W$:
\begin{eqnarray}
    \label{eqn:Hbobywigner}
    \hbH_{BO} &=& \mathcal{W}^{-1}\hat{\Lambda}^{BO}_W  \\
    & = & \mathcal{W}^{-1} \left( \hU^{\dagger}_W \hH_W \hU_W 
\right) \nonumber \\ 
& = & \frac{\bP^2}{2M} + \hat{\Lambda}_{BO}(\bR) \nonumber
\end{eqnarray}

Now, clearly, $\hbH_{BO}$ is an approximation to the full $\hbH$. That being said,  note 
that, because $\hU_W$ depends only on $R$, $\mathcal{W}^{-1}  \hU^{\dagger}_W$ is in fact a unitary operator (in the nuclear-electronic world). 
Therefore, since similarity transforms preserve eigenvalues, it is clear that one can obtain an exact representation of the full Hamiltonian by replacing the matrix in Eq. \ref{eqn:Hbobywigner} with star products:
\begin{eqnarray}
    \hbH
    & = & \mathcal{W}^{-1} \left( \hU^{\dagger}_W *\hH_W * \hU_W 
\right)
\label{eq:BO_exact_back}
\end{eqnarray}

At this point, we can plug in the definition of the star product from Eq. \ref{eqn:starp} and perform a tedious calculation:
\begin{eqnarray}
    \hbH
    & = & \mathcal{W}^{-1} \left( \left( \hU^{\dagger}_W \hH_W + \frac{i \hbar}{2} \frac{\partial \hU^{\dagger}_W}{\partial R} 
    \frac{\partial \hH_W}{\partial P} -
    \frac{\hbar^2}{8} \frac{\partial^2 \hU^{\dagger}_W}{\partial R^2} 
    \frac{\partial^2 \hH_W}{\partial P^2}
\right) * \hU_W 
\right) \\
& = & \mathcal{W}^{-1} \left( \left( \hU^{\dagger}_W \hH_W + \frac{i \hbar}{2} \frac{\partial \hU^{\dagger}_W}{\partial R} 
    \frac{P}{M} -
    \frac{\hbar^2}{8} \frac{\partial^2 \hU^{\dagger}_W}{\partial R^2} 
    \frac{1}{M}
\right) * \hU_W 
\right) 
\\
& = & \mathcal{W}^{-1} \Biggl( \left( \hU^{\dagger}_W \hH_W + \frac{i \hbar}{2} \frac{\partial \hU^{\dagger}_W}{\partial R} 
    \frac{P}{M} -
    \frac{\hbar^2}{8} \frac{\partial^2 \hU^{\dagger}_W}{\partial R^2} 
    \frac{1}{M}
\right) \hU_W  + 
\\
& & \; \; \; \; \; \; 
-\frac{i\hbar}{2} 
\left( \hU^{\dagger}_W \frac{\partial \hH_W}{\partial P} + \frac{i \hbar}{2M} \frac{\partial \hU^{\dagger}_W}{\partial R} 
\right) \frac{\partial \hU_W}{\partial R}
\Biggr)
\\
& & \; \; \; \; \; \; 
-\frac{\hbar^2}{8} 
\left( \hU^{\dagger}_W \frac{\partial^2 \hH_W}{\partial P^2} 
\right) \frac{ \partial^2 \hU_W}{\partial R^2}
\Biggr) \\
& = & \mathcal{W}^{-1} \Biggl( \left( \hU^{\dagger}_W \hH_W + \frac{i \hbar}{2} \frac{\partial \hU^{\dagger}_W}{\partial R} 
    \frac{P}{M} -
    \frac{\hbar^2}{8} \frac{\partial^2 \hU^{\dagger}_W}{\partial R^2} 
    \frac{1}{M}
\right) \hU_W  + 
\\
& & \; \; \; \; \; \;  
 -\frac{i \hbar}{2} \hU^{\dagger}_W \frac{\partial \hU_W}{\partial R} \frac{P}{M} + \frac{ \hbar^2}{4M} \frac{\partial \hU^{\dagger}_W}{\partial R} \frac{\partial \hU_W}{\partial R}
\\
& & \; \; \; \; \; \; 
-\frac{\hbar^2}{8} 
 \hU^{\dagger}_W 
 \frac{ \partial^2 \hU_W}{\partial R^2} \frac{1}{M}
\Biggr) \\
& = & \mathcal{W}^{-1} \Biggl( \hU^{\dagger}_W \hH_W  \hU_W  + \frac{i \hbar}{2}
\left(
\frac{\partial \hU^{\dagger}_W}{\partial R}  \hU_W 
 -  
 \hU^{\dagger}_W \frac{\partial \hU_W}{\partial R} \right) \frac{P}{M} 
 \\
& & \; \; \; \; \; \; 
 + \frac{ \hbar^2}{4M} \frac{\partial \hU^{\dagger}_W}{\partial R} \frac{\partial \hU_W}{\partial R}
 -
    \frac{\hbar^2}{8M} \frac{\partial^2 \hU^{\dagger}_W}   {\partial R^2} 
     \hU_W 
-\frac{\hbar^2}{8M} 
 \hU^{\dagger}_W 
 \frac{ \partial^2 \hU_W}{\partial R^2} 
\Biggr) \\
& = & \mathcal{W}^{-1} \Biggl( \hU^{\dagger}_W \hH_W  \hU_W  + \frac{i \hbar}{2}
\left(
\frac{\partial \hU^{\dagger}_W}{\partial R}  \hU_W 
 -  
 \hU^{\dagger}_W \frac{\partial \hU_W}{\partial R} \right) \frac{P}{M} 
 \\
& & \; \; \; \; \; \; 
 +\frac{ \hbar^2}{4M} \frac{\partial \hU^{\dagger}_W}{\partial R} \frac{\partial \hU_W}{\partial R}
 -
    \frac{\hbar^2}{8M} 
    \frac{\partial}{\partial R }
    \left(
    \frac{\partial \hU^{\dagger}_W}   {\partial R} 
     \hU_W +
 \hU^{\dagger}_W 
 \frac{ \partial \hU_W}{\partial R} \right)
\Biggr) \\
& &   \; \; \; \; \; \; +
\frac{ \hbar^2}{8M} \frac{\partial \hU^{\dagger}_W}{\partial R} \frac{\partial \hU_W}{\partial R}
 +
 \frac{ \hbar^2}{8M} \frac{\partial \hU^{\dagger}_W}{\partial R} \frac{\partial \hU_W}{\partial R}
\\
 & = & \mathcal{W}^{-1} \Biggl( \hU^{\dagger}_W \hH_W  \hU_W  + \frac{i \hbar}{2}
\left(
\frac{\partial \hU^{\dagger}_W}{\partial R}  \hU_W 
 -  
 \hU^{\dagger}_W \frac{\partial \hU_W}{\partial R} \right) \frac{P}{M} 
 \\
& & \; \; \; \; \; \; 
 + \frac{ \hbar^2}{2M} \frac{\partial \hU^{\dagger}_W}{\partial R} \frac{\partial \hU_W}{\partial R}
\end{eqnarray}

The above result can be written succinctly as
\begin{eqnarray}
    \hbH
 & = & \mathcal{W}^{-1} \left( \hU^{\dagger}_W \hH_W  \hU_W  - i \hbar \hat{d}_W \frac{P}{M} 
 + \frac{ \hbar^2}{2M} \hat{\zeta}_W \right)
 \label{eq:finalBO}
\end{eqnarray}

where
\begin{eqnarray} 
\hat{d}_W =   \hU^{\dagger}_W \frac{\partial \hU_W}{\partial R}\\
\hat{\zeta}_W = \frac{\partial \hU^{\dagger}_W}{\partial R} \frac{\partial \hU_W}{\partial R}
\end{eqnarray}

Finally, the end result is:
\begin{eqnarray}
\hbH & = & \frac{\bP^2}{2M} + \hat{\Lambda}_{BO}(\bR) -
\frac{i\hbar}{2M }\left( 
\bP \hat{d}(\bR) + \hat{d}(\bR) \bP \right) 
+\frac{\hbar^2}{2M }\hat{\zeta}(\bR)
\label{eq:finalBOW}
\end{eqnarray} 
where
\begin{eqnarray} 
d_{nj} = \left<\Phi_n \middle | \frac{\partial }{\partial R}  \Phi_j\right> \\
\zeta_{nj} = \left< \frac{\partial}{\partial R} \Phi_n \middle | \frac{\partial}{\partial R} \Phi_j\right>
\end{eqnarray}

Note that Eq. \ref{eq:finalBOW} is identical to Eq. \ref{eq:stdbo:1}. These are both exact transformations of the full Hamiltonian with the same $\hbar$ expansion.  

Lastly, let us consider the practical implications of this expansion for a system without spin  (and time reversal symmetry).  We denote the ground state as level 0, and we seek to quantify the  ground state vibrational energies.

\begin{itemize}
\item The zeroth order BO vibrational energies arise from diagonalizing $\hbH_{BO}$,
\begin{eqnarray}
    \hbH_{BO}^{(0)} \psi = \left(\frac{\bP^2}{2M} + \hat{\Lambda}_{BO}(\bR)\right)\psi = E_{BO}^{(0)} \psi 
\end{eqnarray}

\item Because $d_{00}=0$, there is no first order BO vibrational correction unless we are prepared to mix the ground state with excited states.

\item 
The simplest second order BO vibrational energies arise
by including the diagonal $\zeta_{00}$ correction, i.e. the so-called diagonal BO correction (DBOC): 
\begin{eqnarray}
[V_{\mathrm{DBOC}}]_{00}(\bR) &=& \zeta_{00}(\bR) = \sum_j \frac{d_{0j}d_{j0}}{2M}
\\
\label{eqn:Hbo2}
   \left(\hbH_{BO}^{(0)} +  V_{DBOC} \right) \psi &=& \left(\frac{\bP^2}{2M} + \hat{\Lambda}_{BO}(\bR) + V_{DBOC}\right) \psi = \tilde{E}_{BO}^{(2)} \psi 
\end{eqnarray}
\end{itemize}

To go beyond these expressions, we must mix the ground and excited electronic states, where we will find other second order terms.

\subsection{The Littlejohn-Flynn Perturbative Expansion}
Interestingly,
 Eq. \ref{eq:finalBO} is not the only possible $\hbar$ expansion for the total Hamiltonian that starts with the BO approximation as the zeroth order Hamiltonian. In particular, suppose one starts with the BO electronic states characterized by the electronic transformation $\hU_W$. Rather than recovering the exact nuclear-electronic Hamiltonian by diagonalizing the entire Hamiltonian in Eq. \ref{eq:stdbo:1},  Littlejohn and Flynn showed that one can use perturbation theory and formally diagonalize the total Hamiltonian operator through an 
 infinite sum \cite{littlejohn:flynn:1991:pra:coriolis}:
 \begin{eqnarray}
 \label{eq:hlf:expansion}
\hbH_{LF} = 
\hbH_{LF}^{(0)} + 
\hbH_{LF}^{(1)} + \hbH_{LF}^{(2)} + \ldots
\end{eqnarray}
 The trick is to  perturb the 
electronic transformation $\hU_W$ in a prescribed fashion that incorporates dynamical effects:
\begin{eqnarray}
    \hK_W = \hU_W + \hbar \hU_W^{(1)} + \hbar^2 \hU_W^{(2)} + \ldots
    \label{eqn:defK}
\end{eqnarray}
There are two constraints on the operator $\hK_W$ we seek.

\begin{enumerate}

\item We want $\hK_W$ to be unitary, so that we insist that:

\begin{eqnarray}
\hK_W^{\dagger} * \hK_W &=& \hat{I}     
\end{eqnarray}

\item We want to diagonalize the matrix $\hbH$, so we insist that 
\begin{eqnarray}
\hbH_{LF} = \mathcal{W}^{-1} \left(\hK_W^{\dagger} * \hH_W * \hK_W\right)
\label{eqn:defHLF}
\end{eqnarray}
is diagonal.  
\end{enumerate}

Let us now investigate these constraints order by order in $\hbar$.

\begin{itemize}
\item To zeroth order, these constraints are already satisfied:
\begin{eqnarray}
\hU_W^{\dagger}   \hU_W  &=& \hat{I}  \\
\hU_W^{\dagger}  \hH_W  \hU_W & & \mbox{is diagonal}
\end{eqnarray}

\item To first order, the two constraints determine $\hU_W^{(1)}$:
\begin{eqnarray}
\label{eqn:HLFsolvefirst}
   \left( \hU_W^{(1)}\right)^{\dagger}    \hU_W  +  \hU_W^{\dagger}    \hU_W^{(1)} & = & - \left\{ \hU_W^{\dagger}, \hU_W \right\} = 0 
\end{eqnarray}
\begin{eqnarray}
\label{eqn:HLFsolvesecond}
\left( \hU_W^{(1)}\right)^{\dagger}  \hH_W  \hU_W  +  \hU_W^{\dagger}  \hH_W  \hU_W^{(1)} 
  - i  \hat{d}_W \frac{P}{M} & & \mbox{must be diagonal}
\end{eqnarray}
Note that the right hand side of Eq. \ref{eqn:HLFsolvefirst} is zero because $\hU_W$ depends only on $R$ (not $P$).  From this same equation, it follows that $\hU_W^{\dagger}    \hU_W^{(1)} $ is anti-hermitian, i.e. that 
\begin{eqnarray}
    \hU_W^{\dagger}    \hU_W^{(1)} = \hA,
\end{eqnarray}
where $\hA$ is anti-hermitian. Now, 
inverting this relation
\begin{equation}
    \hU_W^{(1)}=\hU_W \hA 
\end{equation}
and plugging into Eq. \ref{eqn:HLFsolvesecond}, we find that
\begin{eqnarray}
     \hY &\equiv& -\hA \hU_W^{\dagger} \hH_W  \hU_W + \hU_W^{\dagger} \hH_W  \hU_W \hA - i  \hat{d}_W \frac{P}{M} \\
    &=&  -\hA \hLambda_W +  \hLambda_W \hA- i  \hat{d}_W \frac{P}{M} 
\end{eqnarray}
must be diagonal.
In other words, for $k \neq j$,
\begin{eqnarray}
    Y_{kj} = -A_{kj} (\Lambda_W)_{jj} + (\Lambda_W)_{kk} A_{kj} - i  (d_W)_{kj} \frac{P}{M} =0 
\end{eqnarray}
so that the solution is 
\begin{eqnarray}
    A_{kj} = \frac{i  (d_W)_{kj} \frac{P}{M}}{(\Lambda_W)_{kk}-(\Lambda_W)_{jj}}
\end{eqnarray}
and
\begin{eqnarray}
\label{eqn:lfUw1}
    \left(U_W^{(1)}\right)_{nj}
 & = & \sum_{k \ne j} i (U_W)_{nk} \frac{  \left( d_W \right)_{kj} \frac{P}{M}}{\lambda_{kk}^{BO} - \lambda_{jj}^{BO} } =  -\sum_{k \ne j}i (U_W)_{nk}\frac{  \left( \frac{\partial H_W}{\partial R} \right)_{kj} \frac{P}{M}}{ (\lambda_{kk}^{BO} - \lambda_{jj}^{BO})^2 } 
\end{eqnarray}
\end{itemize}


At this point, given the transformation in Eq. \ref{eqn:defK} and the definition in Eq. \ref{eqn:defHLF}, we can write a different (but still exact) expansion of the total Hamiltonian:
\begin{eqnarray}
\label{eq:hlf0}
    \hbH_{LF}^{(0)} = & &  \mW^{-1} \left(  \hU_W^{\dagger} \hH_W \hU_W \right) = \hbH_{BO}^{(0)}
\\
\label{eq:hlf1}
    \hbH_{LF}^{(1)} = & & \mW^{-1} \Biggl(  \hbar (\hU_{W}^{(1)})^{\dagger} \hH_W \hU_W + \hbar \hU_W^{\dagger} \hH_W \hU_W^{(1)} \nonumber \\
    & & + 
    \frac{i \hbar}{2}\left\{ \hU_W^{\dagger}, \hH_W \right\} \hU_W  +     \frac{i \hbar}{2} \left\{\hU_W^{\dagger}  \hH_W,\hU_W \right\} \Biggr) \nonumber \\
 = & & \mW^{-1}\Biggl(  \hbar (\hU_{W}^{(1)})^{\dagger} \hH_W \hU_W + \hbar \hU_W^{\dagger} \hH_W \hU_W^{(1)}   \nonumber \\
    & & + 
    \frac{i \hbar}{2}\left\{ \hU_W^{\dagger}, \hH_W \right\} \hU_W +     \frac{i \hbar}{2} \hU_W^{\dagger} \left\{  \hH_W,\hU_W \right\}  
    \Biggr) \nonumber \\
 = & & \mW^{-1}\Biggl( \hbar (\hU_{W}^{(1)})^{\dagger} \hH_W \hU_W + \hbar \hU_W^{\dagger} \hH_W \hU_W^{(1)}   - i  \hat{d}_W \frac{P}{M}   
   \Biggr)   \\
\label{eq:hlf2}
\hbH_{LF}^{(2)} = & &  
    \mW^{-1}\Biggl(\hbar^2  (\hU_W^{(1)})^{\dagger} \hH_W \hU_W^{(1)} + \hbar^2 (\hU_W^{(2)})^{\dagger} \hH_W \hU_W + \hbar^2 \hU_W^{\dagger} \hH_W \hU_W^{(2)}
    \nonumber
    \\
    & & + 
    \frac{i \hbar^2}{2}\left\{ \hU_W^{\dagger}, \hH_W \right\} \hU_W^{(1)}+     \frac{i \hbar^2}{2}\left\{ \hU_W^{\dagger} \hH_W,\hU_W^{(1)} \right\} 
         \nonumber \\
        & & + 
  \frac{i \hbar^2}{2}\left\{
  (\hU_W^{(1)})^{\dagger}, \hH_W\right\} \hU_W +     \frac{i \hbar^2}{2}\left\{ (\hU_W^{(1)})^{\dagger} \hH_W,\hU_W \right\}     \nonumber 
  \\
& & \frac{- \hbar^2}{4} \left\{ \left\{  \hU_W^{\dagger}, \hH_W \right\}, \hU_W \right\}   -\frac{ \hbar^2}{8} \left\{ \left\{ \hU_W^{\dagger}, \hH_W \right\} \right\}\hU_W \nonumber 
\\
& & -\frac{ \hbar^2}{8}\left\{ \left\{ \hU_W^{\dagger}\hH_W, \hU_W \right\} \right\}\ \Biggl)
\end{eqnarray}

Here, we have defined the ``double Poisson bracket'' to be:
\begin{equation}
    \left\{ \left\{ \hat{O_1} , \hat{O_2}\right\} \right\}\ = (\nabla^2_R \hat{O_1})( \nabla^2_P \hat{O_2}) +(\nabla^2_P \hat{O_1})( \nabla^2_R \hat{O_2}) - 2(\nabla^2_{RP} \hat{O_1})( \nabla^2_{RP} \hat{O_2})
\end{equation}

Let us now revisit the vibrational energies on the ground electronic state within the Littlejohn-Flynn expansion.   At zeroth order, the Littlejohn-Flynn vibrational energies are BO vibrational energies. At first order,  note that $\left( \hat{\bH}_{LF}^{(1)} \right)_{00}$ = 0. To prove this equality, note that:
\begin{eqnarray}
    & &\left( \hbar (\hU_{W}^{(1)})^{\dagger} \hH_W \hU_W + \hbar \hU_W^{\dagger} \hH_W \hU_W^{(1)}   - i  \hat{d}_W \frac{P}{M} \right) _{00} \\
     = & &-A_{00} (\Lambda_W)_{00} + (\Lambda_W)_{00} A_{00} - i  (d_W)_{00} \frac{P}{M}  \\ 
     = & & - i  (d_W)_{00} \frac{P}{M}
\end{eqnarray}
Thus, just as for the usual BO expansion, there is no first order diagonal correction to the Hamiltonian within Littlejohn-Flynn theory; within this perturbative series, all corrections to standard BO theory occur at second and higher order.

\subsection{Summary}
In summary, whether working through the usual BO approach or the perturbative LF expansion, there is only one unique zeroth order vibrational energy and no first order vibrational energy correction. As far as second order results are concerned, we will work with the simple DBOC operator from  Eq. \ref{eqn:Hbo2}:

\begin{table}[h]
\centering
\begin{tabular}{|c|c|}
\hline
method & definition \\
\hline
$E_{BO}^{(0)}$         &  Eq. \ref{eqn:HBO}\\
\hline
$\tilde{E}_{BO}^{(2)}$        &  Eq. \ref{eqn:Hbo2} \\
\hline
\end{tabular}
\caption{Different BO-based energies compared below.}
\end{table}


\section{A Phase Space Electronic Structure Approach through a Wigner-Weyl Transformation}
\label{sec:ps}
All of the theory above might have appeared to be unnecessary so far.  After all, one can work in the BO representation without Wigner-Weyl transforms; and unless one seeks second order energies, Littlejohn-Flynn theory is equivalent to BO theory.
Thus, one might wonder why did we work out all of the calculations above?  

The answer, of course, is that for PS theory, where we parameterize electronic states by both nuclear position $R$ and nuclear momentum $P$, one is forced to use Wigner-Weyl transforms. Moreover, the LF perturbative expansion above can be easily adapted to a PS reference so as to extract improved eigenvalues, and in this scenario, we {\bf do} find first order corrections.

Let us  begin by defining the phase space electronic Hamiltonian 
\begin{eqnarray}
    \hH^{PS2}_W  &\equiv& \frac{(P - i\hbar \hGamma)^2}{2M}  + \hH_{el} \\ & = & 
    \hH_W - \frac{i\hbar \hGamma P}{M} - \frac{\hbar^2 \hGamma^2 }{2M}
\end{eqnarray}
For now, the $\Gamma$ operator can be any electronic operator parametrized by nuclear position $R$, $\hGamma = \hGamma(R)$.  Intuitively, one can consider $\hH^{PS2}_W$ to be the Legendre transform of $\hH_{W}$ if one thinks of $\hGamma$ as ``$\dot{R}$''. Next, just as the case for BO theory, we diagonalize $\hH^{PS}_W$:
\begin{eqnarray}
\label{eq:lambdaps2}
    \hH^{PS2}_W = \hL_W \hLambda^{PS2}_W  \hL_W^{\dagger}
\end{eqnarray}

Note that we can also ignore the $\hGamma^{2}$ term and define
\begin{eqnarray}
    \hat{H}^{PS1}_W  \equiv  \hH_W - \frac{i\hbar \hGamma P}{M} 
\end{eqnarray}
which is then diagonalized:
\begin{eqnarray}
    \hH^{PS1}_W = \hL_W \hLambda^{PS1}_W  \hL_W^{\dagger}
    \label{eq:lambdaps1}
\end{eqnarray}

\subsection{A perturbative expansion for a phase space electronic Hamiltonian}
Now, in the spirit of Eq. \ref{eq:BO_exact_back}, one would like to generate an exact representation of the total Hamiltonian using the PS eigenvectors:

\begin{eqnarray}
\label{eqn:FPSexpansion}
    \hbH_{FPS}
        & \stackrel{?}{=} & \mathcal{W}^{-1} \left( \hL^{\dagger}_W * \hH_W  * \hL_W 
\right) \\ 
    & \stackrel{?}{=} & \mathcal{W}^{-1} \left( \hL^{\dagger}_W *( \hH^{PS}_W  + i\hbar \frac{\hGamma P}{M} + \frac{\hbar^2 \hGamma^2 }{2M}) * \hL_W 
    \right) \nonumber
\end{eqnarray}
and develop a corresponding expansion:

\begin{eqnarray}
    \hbH_{FPS} = & &
    \hbH_{FPS}^{(0)} +
        \hbH_{FPS}^{(1)} +
            \hbH_{FPS}^{(2)} +
                \hbH_{FPS}^{(3)} + \ldots 
    \label{eq:PS:expand_in_hbar}
\end{eqnarray}

However, one cannot map Eq. \ref{eqn:FPSexpansion} to Eq. \ref{eq:BO_exact_back} because $\mathcal{W}^{-1} \hL^{\dagger}_W $ is not  unitary in the full nuclear-electronic Hilbert space.
Instead, one must imagine perturbing the $\hL^{\dagger}_W$ operator to find a nearby unitary operator (here, denoted as $\hJ_W$):
\begin{eqnarray}
\hJ_W =  \hL_W 
+ \hbar \hL^{(1)}_W 
+ \hbar^2 \hL^{(2)}_W 
+ \hbar^3 \hL^{(3)}_W + \ldots
\end{eqnarray}

Thereafter, one can construct the series in Eq. \ref{eq:PS:expand_in_hbar} by matching terms with Eq. \ref{eq:exact_ps}.
\begin{eqnarray}
\label{eq:exact_ps}
    \hbH
        & = & \mathcal{W}^{-1} \left( \hJ_W * \hH_W  * \hJ^{\dagger}_W 
\right) 
\\ \nonumber
    & = & \mathcal{W}^{-1} \Bigg[ \left(\hL_W 
+ \hbar \hL^{(1)}_W  
+ \hbar^2 \hL^{(2)}_W 
+ \hbar^3 \hL^{(3)}_W + \ldots \right) *( \hH^{PS}_W  + i\hbar \frac{\hGamma P}{M} + \frac{\hbar^2 \hGamma^2 }{2M}) \\ \nonumber
& &  * \left(\hL_W 
+ \hbar \hL^{(1)}_W 
+ \hbar^2 \hL^{(2)}_W
+ \hbar^3 \hL^{(3)}_W + \ldots \right)^{\dagger}
    \Bigg] 
\end{eqnarray}

\subsubsection{Enforcing Unitarity}
If we are to enforce that $\hJ = \mathcal{W}^{-1} \hJ_W$  be unitary, mathematically we require that:
\begin{eqnarray}
\hJ_W^{\dagger} * \hJ_W = \hat{I}
\end{eqnarray}
If we go order by order in $\hbar$, the relevant conditions are:

\begin{itemize}
    \item For $\mathcal{O}(\hbar)$, $\hL_W^{(1)}$ must satisfy:
\begin{eqnarray}
\label{eq:solvehbar1}
    \frac{i \hbar}{2} \left\{  \hL_W^{\dagger}, \hL_W \right\} + \hbar \left( (\hL_W^{(1)})^{\dagger} \hL_W  + \hL_W^{\dagger}\hL_W^{(1)} \right)  = 0
\end{eqnarray}

\item For order $\mathcal{O}(\hbar^2)$, $\hL_W^{(2)}$ must satisfy:
\begin{eqnarray}
    \label{eq:solvehbar2}
         \frac{- \hbar^2}{8} \left\{ \left\{  \hL_W^{\dagger}, \hL_W \right\} \right\} + \hbar^2 \left( (\hL_W^{(2)})^{\dagger}\hL_W  + \hL_W^{\dagger}\hL_W^{(2)} + (\hL_W^{(1)})^{\dagger} \hL_W^{(1)} \right)  \\
         +  \frac{i\hbar^2}{2} \left( {\left\{  \hL_W^{\dagger}, \hL_W^{(1)} \right\}+
         \left\{   (\hL_W^{(1)})^{\dagger}, \hL_W \right\}} \right)= 0 \nonumber
\end{eqnarray}
\end{itemize}

Here, we work to first order only. 
The solution to Eq. \ref{eq:solvehbar1} is:
\begin{eqnarray}
\label{eq:param:lA}
    \hL_W^{(1)} =      \hL_W \left( -\frac{i}{4} \left\{  \hL_W^{\dagger}, \hL_W \right\} + \hA\right)
\end{eqnarray}
where $\hA$ is any antisymmetric electronic matrix.

\subsubsection{Diagonalizing the Hamiltonian}
In order to specify $\hA$, we will further require that the matrix
\begin{eqnarray}
\hH^W_{FPS} = \hJ_W^{\dagger} * \hH_W * \hJ_W 
\end{eqnarray}
be block diagonal (with respect to the electronic components).
If we go order by order in $\hbar$, the conditions are as follows.
\begin{itemize}
\item To zeroth order,  the operator
\begin{eqnarray}
\label{eq:hps0}
\hH_{FPS}^{(0) W} =   \hL_W^{\dagger} \hH^{PS}_W \hL_W
\end{eqnarray}
must be diagonal. This constraint is obviously satisfied in a basis of phase space eigenvectors (see Eq. \ref{eq:lambdaps2}). 

\item  To first order in  $\mathcal{O}(\hbar)$, we require that
\begin{eqnarray}
\begin{aligned}
\label{eq:fps1}
 \hH_{FPS}^{(1) W} = \hbar (\hL_{W}^{(1)})^{\dagger} \hH_W^{PS} \hL_W + \hbar \hL_W^{\dagger} \hH_W^{PS} \hL_W^{(1)} + i\hbar \frac{P}{M} \left( \hL_W^{\dagger} \hGamma  \hL_W \right)  \\
     + 
    \frac{i \hbar}{2}\left\{ \hL_W^{\dagger}, \hH_W^{PS} \right\} \hL_W +     \frac{i \hbar}{2} \hL_W^{\dagger} \left\{  \hH_W^{PS},\hL_W \right\}  \\
     +
    \frac{i \hbar}{2} \frac{\partial \hL_W^{\dagger}} {\partial R} \hH_W^{PS}
\frac{\partial \hL_W}{\partial P} 
-
    \frac{i \hbar}{2} \frac{\partial \hL_W^{\dagger}} {\partial P} \hH_W^{PS}
\frac{\partial \hL_W}{\partial R} 
\end{aligned}
\end{eqnarray}
must be diagonal.
In short hand form, 
\begin{eqnarray}
    \hbar \hC   + \hbar \left( (\hL_W^{(1)})^{\dagger} \hH_W^{PS} \hL_W  + \hL_W^{\dagger} \hH_W^{PS}\hL_W^{(1)} \right)  
\end{eqnarray}
\noindent must be diagonal where
\begin{eqnarray}
    \hC & =  i \frac{P}{M} \left( \hL_W^{\dagger} \hGamma  \hL_W \right) \nonumber +
    \frac{i}{2}\left\{ \hL_W^{\dagger}, \hH_W^{PS} \right\} \hL_W +     \frac{i}{2} \hL_W^{\dagger} \left\{  \hH_W^{PS},\hL_W \right\}  \\
     & +\frac{i}{2} \frac{\partial \hL_W^{\dagger}} {\partial R} \hH_W^{PS}\frac{\partial \hL_W}{\partial P} -\frac{i}{2} \frac{\partial \hL_W^{\dagger}} {\partial P} \hH_W^{PS}
\frac{\partial \hL_W}{\partial R} 
\end{eqnarray}
Plugging in the form of $\hL^{(1)}_W$ from Eq. \ref{eq:param:lA}, we find that
\begin{eqnarray}
    \hC   +  \left( 
    \left( -\frac{i}{4} \left\{  \hL_W^{\dagger}, \hL_W \right\} - \hA\right) 
    \hLambda_W^{PS}  + \hLambda_W^{PS} \left( -\frac{i}{4} \left\{  \hL_W^{\dagger}, \hL_W \right\} + \hA\right) \right)  
\end{eqnarray}
muts be diagonal. In other words, 
\begin{eqnarray}
\label{eq:hfps1}
     \hH_{FPS}^{(1) W} = \hbar(\hLambda_W^{PS} \hA - \hA \hLambda_W^{PS} + \hB)
\end{eqnarray}
must be diagonal, where
\begin{eqnarray}
   \hat{B}=  & & i \frac{P}{M} \left( \hL_W^{\dagger} \hGamma  \hL_W \right) \nonumber 
     + 
    \frac{i }{2}\left\{ \hL_W^{\dagger}, \hH_W^{PS} \right\} \hL_W +     \frac{i }{2} \hL_W^{\dagger} \left\{  \hH_W^{PS},\hL_W \right\}  
     +
    \frac{i }{2} \frac{\partial \hL_W^{\dagger}} {\partial R} \hH_W^{PS}
\frac{\partial \hL_W}{\partial P} \\
 & &  -
    \frac{i }{2} \frac{\partial \hL_W^{\dagger}} {\partial P} \hH_W^{PS}
\frac{\partial \hL_W}{\partial R}  
   - 
   \frac{i}{4} \left\{  \hL_W^{\dagger}, \hL_W \right\} \hLambda_W^{PS} -  \frac{i}{4} \hLambda_W^{PS}  \left\{  \hL_W^{\dagger}, \hL_W \right\} 
\end{eqnarray}
We can enforce diagonalizability for Eq. \ref{eq:hfps1} by setting:
\begin{eqnarray}
    A_{ij}= -\frac{B_{ij}}{(\Lambda^{PS}_W)_{ii} - (\Lambda^{PS}_W)_{jj}}
\end{eqnarray}

Thus, the final form for $L_{W}^{(1)}$ (analogous to $U_{W}^{(1)}$ in Eq. \ref{eqn:lfUw1} for LF) is:
\begin{equation}
\begin{gathered}
\label{eqn:L_w1ij}
\left(\hL_W^{(1)}\right)_{n j}=-\frac{i}{4} \sum_{k \neq j}\left(\hL_W\right)_{n k}\left\{\hL_W^{+}, \hL_W\right\}_{k j}-i \sum_{k \neq j}\left(\hL_W\right)_{n k} \frac{\left( \frac{P}{M} \hL_W^{+} \hGamma \hL_W\right)_{k j}}{(\Lambda^{PS}_W)_{kk}-(\Lambda^{PS}_W)_{jj}} \\
-\frac{i}{4} \sum_{k \neq j}\left(\hL_W\right)_{n k} \frac{\left(2\left\{\hL_W^{+}, \hH_W^{PS}\right\} \hL_W+2\left\{\hL_W^{+} \hH_W^{PS}, \hL_W\right\}-\left\{\hL_W^{+}, \hL_W\right\} \hLambda^{PS}_W-\hLambda^{PS}_W\left\{\hL_W^{+}, \hL_W\right\}\right)_{k j}}{(\Lambda^{PS}_W)_{kk}-(\Lambda^{PS}_W)_{jj}} 
\end{gathered}
\end{equation}

\item Just for  the sake of completeness, we note that, to second order, the operator
\begin{eqnarray}
\label{eq:hfps2}
\hH_{FPS}^{(2) W} & =
    \hbar^2  (\hL_W^{(1)})^{\dagger} \hH_W^{PS} \hL_W^{(1)} + \hbar^2 (\hL_W^{(2)})^{\dagger} \hH_W^{PS} \hL_W + \hbar^2 \hL_W^{\dagger} \hH_W^{PS} \hL_W^{(2)}
    \nonumber
    \\
     & + i \hbar^2 (\hL_W^{(1)})^{\dagger} \hGamma \frac{P}{M} \hL_W +  i \hbar^2
    \hL_W^{\dagger} \hGamma \frac{P}{M} \hL_W^{(1)}
        \nonumber \\
    & + 
    \frac{i \hbar^2}{2}\left\{ \hL_W^{\dagger}, \hH_W^{PS} \right\} \hL_W^{(1)}+     \frac{i \hbar^2}{2}\left\{ \hL_W^{\dagger} \hH_W^{PS},\hL_W^{(1)} \right\} 
         \nonumber \\
         & + 
  \frac{i \hbar^2}{2}\left\{
  (\hL_W^{(1)})^{\dagger}, \hH_W^{PS}\right\} \hL_W +     \frac{i \hbar^2}{2}\left\{ (\hL_W^{(1)})^{\dagger} \hH_W^{PS},\hL_W \right\}     \nonumber 
  \\
 & - \frac{ \hbar^2}{2}\left\{ \hL_W^{\dagger}, \hGamma \frac{P}{M} \right\} \hL_W -    \frac{ \hbar^2}{2}\left\{ \hL_W^{\dagger} \hGamma \frac{P}{M} ,\hL_W \right\} 
    \nonumber \\
 & \frac{- \hbar^2}{4} \left\{ \left\{  \hL_W^{\dagger}, \hH_W^{PS} \right\}, \hL_W \right\} -\frac{ \hbar^2}{8}\left\{ \left\{ \hL_W^{\dagger}\hH_W^{PS}, \hL_W \right\} \right\}\
\nonumber \\
 & -\frac{ \hbar^2}{8} \left\{ \left\{ \hL_W^{\dagger}, \hH_W^{PS} \right\} \right\}\hL_W +\hbar^2  \frac{\hL_W^{\dagger} \hGamma^{2} \hL_W}{2M}
\end{eqnarray}
should be diagonal. Though this equation is simple to solve (and in fact identical in form to Eq. \ref{eq:fps1} above), we will not solve Eq. \ref{eq:hfps2} as we work only to first order below.  
\end{itemize}

\subsection{Evaluating the First Order Correction In Practice}
For the phase-space approach, unlike the BO-Littlejohn-Flynn approach, there is a nonzero first order energy correction. To find this correction, let us evaluate $(\hH_{FPS}^{(1) W})_{00}$ in Eq. \ref{eq:fps1}.  Note that $\Lambda_W^{PS}$ is diagonal, so that
\begin{eqnarray}
    (\hLambda_W^{PS} \hA - \hA \hLambda_W^{PS})_{00} = (\Lambda_W^{PS})_{00} A_{00}  - A_{00}(\Lambda_W^{PS})_{00}=0
\end{eqnarray}
Therefore,  
\begin{eqnarray}
(\hbH_{FPS}^{(1) W})_{00} 
= & &  \hbar B_{00} \nonumber \\
= & &  
\hbar\left[  i \frac{P}{M} \left( \hL_W^{\dagger} \hGamma  \hL_W \right) 
+ \frac{i}{2} \left\{ \hL_W^{\dagger}, \hH_W^{PS} \right\} \hL_W 
+ \frac{i}{2} \hL_W^{\dagger} \left\{ \hH_W^{PS}, \hL_W \right\}  \right. \nonumber \\
 & & +\frac{i}{2} \frac{\partial \hL_W^{\dagger}}{\partial R} \hH_W^{PS} \frac{\partial \hL_W}{\partial P} 
- \frac{i}{2} \frac{\partial \hL_W^{\dagger}}{\partial P} \hH_W^{PS} \frac{\partial \hL_W}{\partial R}  \nonumber \\
 & & \left. - \frac{i}{4} \left\{ \hL_W^{\dagger}, \hL_W \right\} \hLambda_W^{PS}
- \frac{i}{4} \hLambda_W^{PS} 
\left\{\hL_W^{\dagger}, \hL_W\right\} \right] _{00}
\label{eq:B00}
\end{eqnarray}
and we can  calculate an energetic correction without evaluating $\hL_{W}^{(1)}$ explicitly.

\subsection{Evaluating the Second Order Correction In Practice}

In practice, evaluating a second order correction will be expensive and impractical for most systems. Nevertheless, just as a benchmark, we have 
implemented a very naive second order correction by keeping only the last term in Eq. \ref{eq:hfps2}:
\begin{eqnarray}
(\hH_{FPS}^{(2) W})_{00} & \approx & \hbar^2 Z_{00} \nonumber\\
&=& \hbar^2  \left( \frac{\hL_W^{\dagger} \hGamma^{2} \hL_W}{2M}\right)_{00}
\label{eq:naive2}
\end{eqnarray}

\subsection{Summary}

We will now summarize several different phase space methods, all of which are compared below.

\begin{table}[H]
    \centering
    \begin{tabular}{|c|c|c|}
    \hline
Method & Symbol(R,P)  &  Eq. number(s)  \\
\hline
$\tilde{E}_{PS1}^{(0)}$         &  $(\Lambda^{PS1}_W)_{00}$ &  Eq. \ref{eq:lambdaps1}
\\
\hline
$\tilde{E}_{PS1}^{(1)}$         &  $(\Lambda^{PS1}_W)_{00} + \hbar B_{00}$ &  Eqs. \ref{eq:lambdaps1}, \ref{eq:B00}
\\
  \hline
    $\tilde{E}_{PS2}^{(0)}$         &  $(\Lambda^{PS2}_W)_{00}$   & Eq. \ref{eq:lambdaps2}
\\
  \hline
    $\tilde{E}_{PS2}^{(1)}$         & $(\Lambda^{PS2}_W)_{00} + \hbar B_{00}$  & Eqs. \ref{eq:lambdaps2}, \ref{eq:B00}
\\
  \hline
    $\tilde{E}_{PS2}^{(2)}$         & $(\Lambda^{PS2}_W)_{00} + \hbar B_{00} + \hbar^2 Z_{00}$  & Eqs. \ref{eq:lambdaps2}, \ref{eq:B00},\ref{eq:naive2}
\\
  \hline
    \end{tabular}
    \caption{A list of the relevant PS methods we will report below.  On the left, we list the abbreviation for the energetic approximation. In the middle, we list the symbol that must be Weyl transformed (Eq. \ref{eqn:weyl}) and then diagonalized.  On the right, we list the relevant equations.}
\end{table}

\section{Results}
\label{sec:model}
\subsection{Model}
For our Hamiltonian of choice, we work with a well-known model developed by Borgis\cite{marinica2006generating} and recently studied by Gross et al.\cite{scherrer2017mass}.
The Hamiltonian of the three-particle one-dimension hydrogen bond model can be written as:
\begin{eqnarray} \label{eq:fullH}
   \hbH =  \frac {\hat P_1^2} {2M} + \frac {\hat P_2^2} {2M} + \frac {\hat p_{\rm e}^2} {2m} +  \hat V(\hat R_1, \hat R_2, \hat r_{\rm e}),
\end{eqnarray} 
where we assume the mass of two heavy particles are the same $M_1 = M_2 = M$. By change of coordinates, we can reduce the number of variables. That is to say, let us go to the molecular (nuclei + electron) center of mass frame, where the molecular center of mass (MCM) is defined as:
\begin{eqnarray} 
    \hat R_{\rm MCM} =  \frac {M\hat R_1 + M\hat R_2 + m\hat r_{\rm e}} {2M + m}, 
\end{eqnarray}

We transform the coordinates from $(\hat R_1, \hat R_2, \hat r_{\rm e})$ to MCM coordinates $(\hat R_{\rm MCM}, \hat R, \hat r)$,
where $\hat R$ is the distance between the two oxygen atoms and the light degrees of freedom $\hat r$ indicates the distance between the H atom to the nuclear center of mass.
\begin{equation}
    \hat R =  \hat R_1 - \hat R_2, \quad 
  \hat r =  \hat r_{\rm e} -  \frac{\hat R_1 + \hat R_2}{2}.  
\end{equation}
The Hamiltonian therefore becomes effectively a two dimensional model if we get rid of the molecular center of mass motion:
\begin{eqnarray} \label{eq:modelH}
    \hbH =  \frac {\hat P_{\rm MCM}^2} {2(2M + m)} + \frac {\hat P^2} {2\mu}  + \frac {{\hat {p}}^2} {2m} +  \frac {{\hat {p}}^2} {2(M+M)}
    + \hat V(\hat R, \hat r), 
\end{eqnarray} 
where $\mu = \frac M 2$ is the nuclear reduced mass and the second to last term is the mass polarization term\cite{davis1982mass}, 
and the momentum operators in the new coordinate system are defined as:
\begin{eqnarray} 
    \hat P_1 = \frac {M \hat P_{\rm MCM}} {2M+ m} + \hat P - \frac {\hat p} {2}, \quad  \hat P_2 = \frac {M \hat P_{\rm MCM}} {2M+ m} - \hat P - \frac {\hat p} {2}, \quad \hat p_{\rm e} = \hat  p + \frac {m \hat P_{\rm MCM}} {2M + m}.
\end{eqnarray} 
In the MCM coordinates, the potential energy term does not depend on the $R_{\rm MCM}$ and can be expressed as:
\begin{eqnarray} \label{eq:modelV}
\hat V(\hat R, \hat r) &=& D\left(e^{-2a\left(\frac {\hat R} 2 + \hat r - d\right)} -2e^{-a\left(\frac {\hat R } 2+ \hat r - d\right)} + 1 \right) \nonumber \\
&+& Dc^2\left(e^{-\frac {2a} c\left(\frac {\hat R} 2 - \hat r - d\right)} -2e^{-\frac a c\left(\frac {\hat R} 2 - \hat r - d\right)}\right) + Ae^{-B\hat R} - \frac C {\hat R^6}. 
\end{eqnarray}
The relevant parameters are given in Table.~\ref{tab:parameters}.
Below, in Figs. \ref{fig:abs_vib_womp}-\ref{fig:logre_vib_wmp},  for exact vibrational energies, we will diagonalize the $\hbH$ operator in Eq. \ref{eq:modelH} (setting $\hP_{MCM} = 0$).

\begin{table}[h]
    \centering
    \begin{tabular}{|c|c|c|c|c|c|c|c|}
        \hline
        $D$ & $d$ & $a$ & $c$ & $A$ & $B$ & $C$ \\
        \hline
        $60~\rm kcal/mol$ & $0.95~\textup{\AA}$ & $2.52~\textup{\AA}^{-1}$ & $0.707$ & $2.32 \times 10^5~\rm kcal/mol$ & $3.15~\textup{\AA}^{-1}$ & $2.31 \times 10^4~\rm kcal/mol/\textup{\AA}^{6}$ \\
        \hline
    \end{tabular}
    \caption{Model parameters from Ref.~\citenum{marinica2006}.}
    \label{tab:parameters}
\end{table}


Turning to approximate methods, let us now delineate the BO and PS constructions. 
\begin{itemize}
\item Within BO theory, we generating potential energy surfaces by diagonalizing the operator:
\begin{eqnarray}
    \hH_{el}= \frac {{\hat {p}}^2} {2m} +  \frac {{\hat {p}}^2} {2(M+M)}
    + \hat V(\hat R, \hat r)
\end{eqnarray}
Thereafter, to generate BO vibrational energies, we diagonalize:
\begin{eqnarray}
    \hbH=\frac {\hbP^2} {2\mu}+\hat V_0(\hbR)
\end{eqnarray}

\item 
As far as phase space electronic structure is concerned, for this model Hamiltonian, we must begin by inserting $\hat{\Gamma}$ operators for both nuclei in Eq. \ref{eq:fullH}.  After we remove the center of mass and reduce to one coordinate, following Ref.\citenum{bian:2025:jctc:wigner_vibrations}, the relevant phase space electronic Hamiltonian is of the form:
\begin{eqnarray}
\label{eq:HPS:model}
   \hat H_{\rm PS} (R,P) = \frac {(P- i\hbar \hat \Gamma)^2} {2\mu} + \frac {\hat p^2} {2m} +  \hat V(R , \hat r), 
\end{eqnarray}
where
\begin{eqnarray} 
      \hat \Gamma  = \frac {\hat \Gamma_1 - \hat \Gamma_2} 2.
\end{eqnarray} 
\begin{eqnarray}
    \hat \Gamma_1 =  \frac{1}{2i\hbar}\left( \hat{\theta}_1\hat{ p}  + \hat{p}  \hat{\theta}_1\right), \quad \hat \Gamma_2 =  \frac{1}{2i\hbar}\left( \hat{\theta}_2\hat{p}  + \hat{p}  \hat{\theta}_2\right)
\end{eqnarray}
with 
\begin{eqnarray} \label{eq:theta:used}
    \hat{\theta}_1 = \frac{e^{-|\hat{r}-\frac R 2|^2/\sigma^2}}
    {e^{-|\hat{r} - \frac R 2|^2/\sigma^2} + {e^{-|\hat{r}+\frac R 2|^2/\sigma^2}}},\quad \hat{\theta}_2 = 
    \frac{e^{-|\hat{r} +\frac R 2|^2/\sigma^2}} {e^{-|\hat{r} - \frac R 2|^2/\sigma^2} + {e^{-|\hat{r}+\frac R 2|^2/\sigma^2}}}.
\end{eqnarray}

Finally, recall that in order to generate eigenvalues through a phase space formalism, we must further perform a Weyl transform on $E_W(R,P)$ ($E = \mathcal{W}^{-1}(E_W)$ in Eq. \ref{eqn:weyltransform} above) and then diagonalize the resulting $E(R,R')$. 

For all PS calculations, we have used a grid size of $N_R = 101$ grid points for nuclear coordinate $R \in \left[2, 4\right] \textup{\AA}$ and $N_P = 101$ grid points for nuclear momentum $P \in \left[-\pi /\Delta R + \pi/(N_P\Delta R), ..., \pi /\Delta R - \pi/(N_P\Delta R) \right]$ atomic units (au) based on the conjugate Fourier transform. For electronic coordinates, we choose $N_r = 200$ grid points for the electronic coordinate $r \in \left[-2, 2\right] \textup{\AA}$. Lastly, in Figs. \ref{fig:abs_vib_womp}-\ref{fig:logre_ge_womp}, 
we do not include the so-called mass polarization term from Eq. \ref{eq:modelH}; the mass polarization term is included in Figs. \ref{fig:abs_ge_wmp}-\ref{fig:logre_vib_wmp}.
\end{itemize}

\subsection{First vibrational Energy as a Function of the reduced mass \texorpdfstring{$\mu$}{mu}}

\begin{figure}[H]
\includegraphics[width=0.9\linewidth, height=0.55\linewidth]{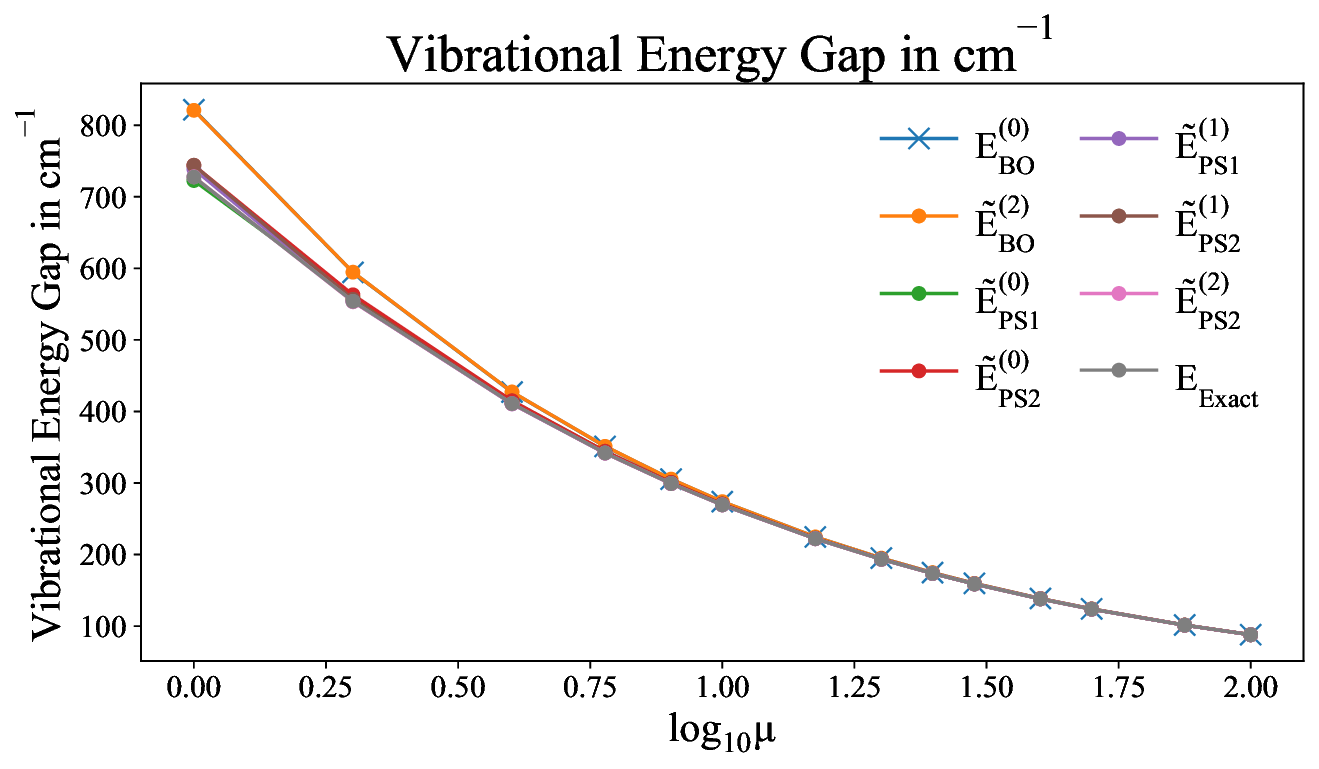}
\caption{\label{fig:abs_vib_womp}  Vibrational energy gap as a function of reduced mass $\mu$ without adding the mass polarization term.  On this energy scale, one cannot discern very much about the quality of different phase space techniques; that being said, the BO wavefunction is clearly the worst performing.}
\end{figure}

\begin{figure}[H]
\includegraphics[width=0.9\linewidth, height=0.55\linewidth]{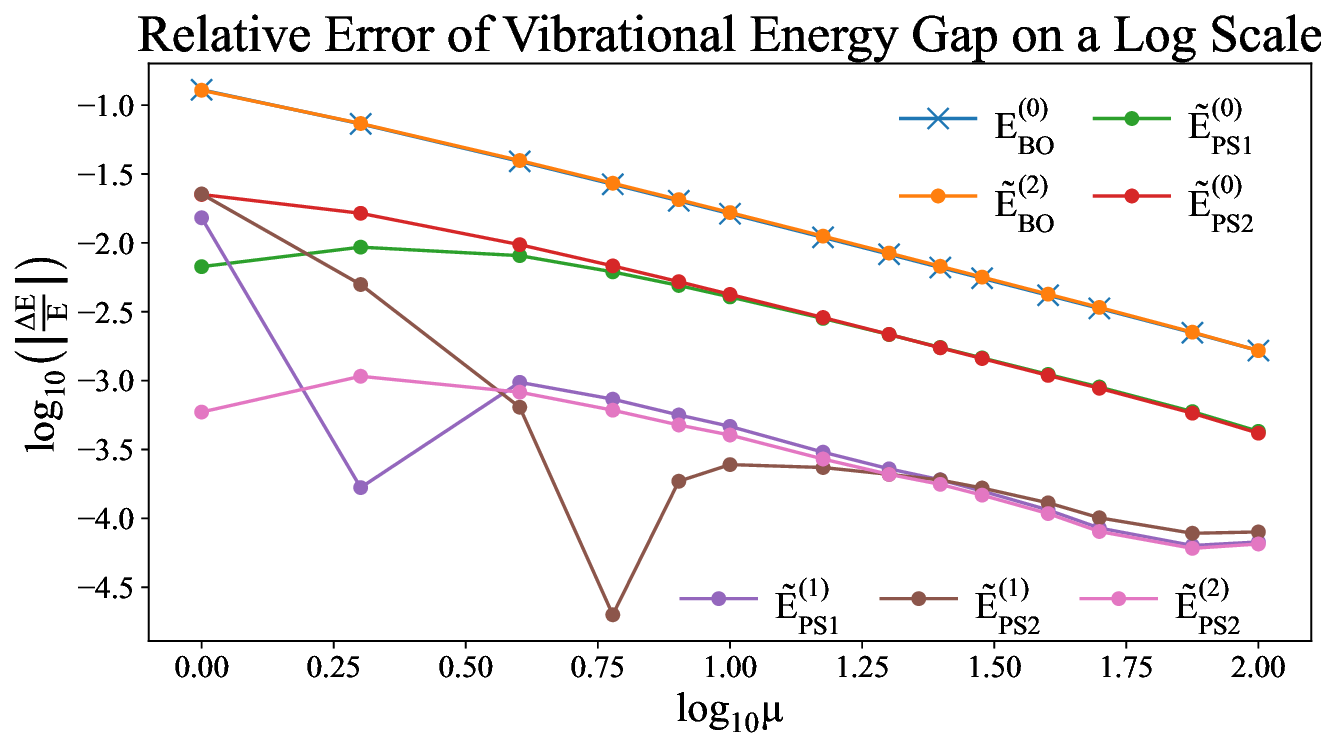}
\caption{\label{fig:logre_vib_womp}  Relative error of the vibrational energy gap as a function of reduced mass $\mu$ on a log scale without adding the mass polarization term.  While zeroth order PS approaches can offer an order of magnitude better accuracy than BO approaches, we can gain another order of magnitude in accuracy by going to higher order in PS theory.  That being said, the $\tilde{E}_{PS1}^{(1)}$ and $\tilde{E}_{PS2}^{(1)}$ energies do not appear to be entirely stable for small $\mu$ values. }
\end{figure}

Our first set of data is reported in Figs. \ref{fig:abs_vib_womp}-\ref{fig:logre_vib_womp}. Here, we find (in agreement with Ref. \citenum{bian:2025:jctc:wigner_vibrations}) that indeed PS vibrational energies are far more accurate than BO vibrational energies. Moreover, one can gain another order of accuracy by correcting phase space methods; in particular, $\tilde{E}_{PS1}^{(1)}$ and $\tilde{E}_{PS2}^{(1)}$ are 100 times more accurate than BO theory. That being said, we must also notice that, when the mass ratio becomes small (small $\mu$), these perturbation theories become less stable and we find a large fluctuation in the final result. Overall, for this set of data, $\tilde{E}_{PS2}^{(2)}$ appears the most stable and accurate. As far as scaling is concerned, note in Fig. \ref{fig:logre_vib_womp}, without the mass polarization term, all the BO and PS errors decay as $\mu^{-1}$, which is the usual BO error. 


\subsection{Harmonic Approximation}
\begin{figure}[H]
\includegraphics[width=0.9\linewidth, height=0.55\linewidth]{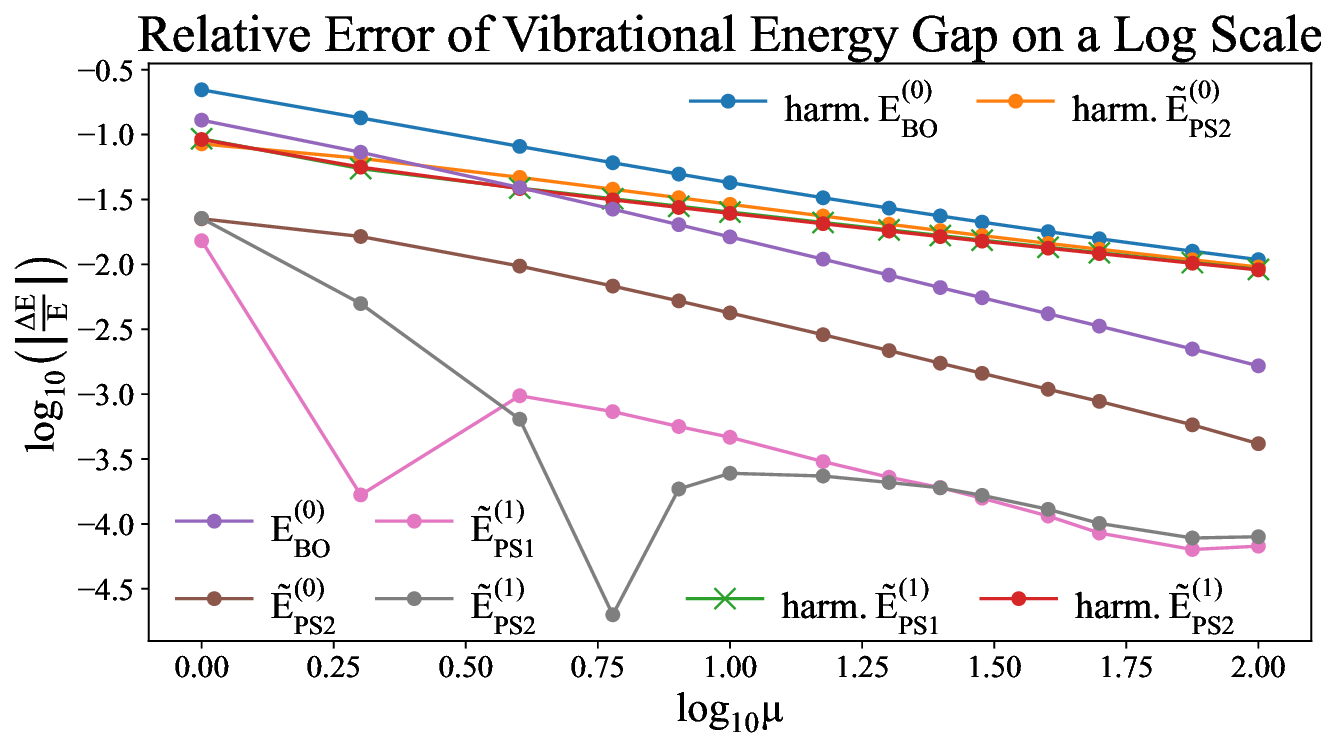}
\caption{\label{fig:logre_ha_vib_womp} Relative error of the vibrational energy gap as a function of reduced mass $\mu$ on a log scale where we now investigate the harmonic limit (without the mass polarization term).  Going beyond the zeroth order phase space energy gives only a modest correction in the harmonic limit.}
\end{figure}

\begin{figure}[H]
\includegraphics[width=0.9\linewidth, height=0.55\linewidth]{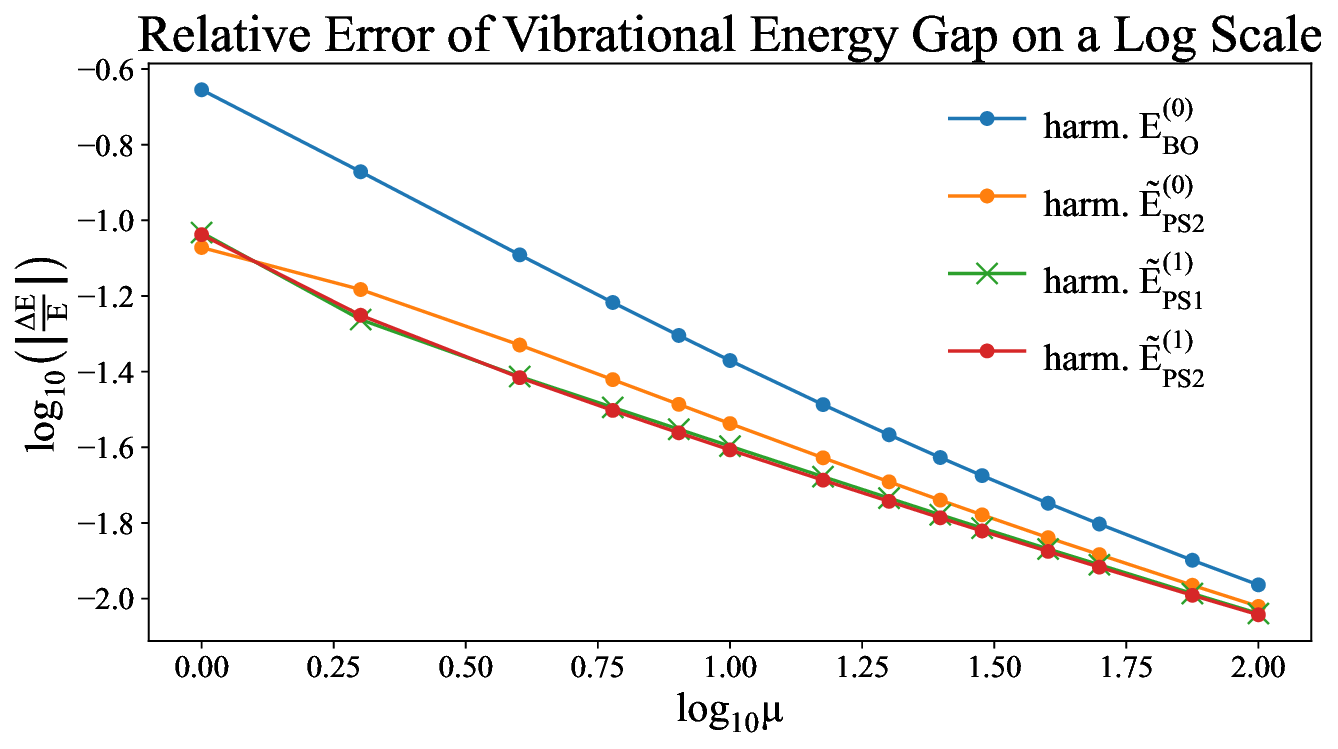}
\caption{\label{fig:zoomin_logre_ha_vib_womp}Zoom in of Fig. \ref{fig:logre_ha_vib_womp} for more readability.}
\end{figure}

For the most complete analysis, we also report here results where we consider harmonic BO and phase-space surfaces, i.e. where we force $E_0(R)$ or $E^{\rm PS}_{W,0}(R,P)$ to be a quadratic function.
For the phase-space surface $E^{\rm PS}_{W,0} (R, P)$, we force the energy to be quadratic with respect to both $P$ and $R$ and expand $E^{\rm PS}_{W,0}$ at the minimum energy point $(R_{\rm min}, 0)$: 
\begin{eqnarray} \label{eq:psharmonic1}
E^{\rm PS, harmonic}_{W, 0} =  \frac{1}{2}\frac {\partial^2 E^{\rm PS}_{W,0}}  {\partial P^2}\bigg|_{P = 0} P^2 + \frac{1}{2} \frac {\partial^2 E^{\rm PS}_{W,0}} {\partial R^2}\bigg|_{R = R_{\rm min}} R^2,
\end{eqnarray}
where the expansion coefficients are calculated by finite difference.
Note that since we have not considered spin-related couplings and there is only one vibrational mode, the phase-space energy will always be minimized at $P = 0$ and the off-diagonal Hessian will be 0, i.e., $\frac {\partial^2 E^{\rm PS}_{W,0}} {\partial R \partial P }= 0$. 
The phase-space harmonic vibrational energy gap is then computed by
\begin{eqnarray} \label{eq:psharmonic2}
    \Delta E = \hbar  \sqrt{\frac {\partial^2 E^{\rm PS}_{W,0}} {\partial R^2} \frac {\partial^2 E^{\rm PS}_{W,0}} {\partial P^2}}.
\end{eqnarray}
According to Fig. \ref{fig:logre_ha_vib_womp}, note that, whereas we can gain a strong correction by using phase space approach rather than BO in the harmonic limit, there is only a marginal gain in accuracy when we correct phase space theory in the harmonic (and move from $\tilde{E}_{PS2}^{(0)}$ to $\tilde{E}_{PS1}^{(1)}$ or $\tilde{E}_{PS2}^{(1)}$).  For all of the harmonic data, the error falls off as $\mu^{-1/2}$, indicating that the insurmountable problem for these calculations is the anharmonic coupling.

\section{Discussion: Mass polarization term and Absolute Ground State Energies}
\label{sec:mpandha}
Thus far, we have not focused on the mass polarization term  defined by
\begin{equation}
    \hH_{mp}=\frac {{\hat {p}}^2} {2(M+M)}
\end{equation}
in Eq. \ref{eq:modelH}. 
For exact diagonalization calculations above, we have included this term. However, within a phase-space approach,  we have not included such a mass polarization term above (for reasons that will soon be clear).
In this discussion section we will now address the question of mass polarization in the context of phase space electronic structure theory.  In short, for the impatient reader, we find that including mass polarization can change absolute ground state energies but has a much more modest affect on vibrational energies (that require only relative ground state energy information). This finding is encouraging because most calculations will include mass polarization if only to  maintain size consistency.

\begin{figure}[H]
\includegraphics[width=0.9\linewidth, height=0.55\linewidth]{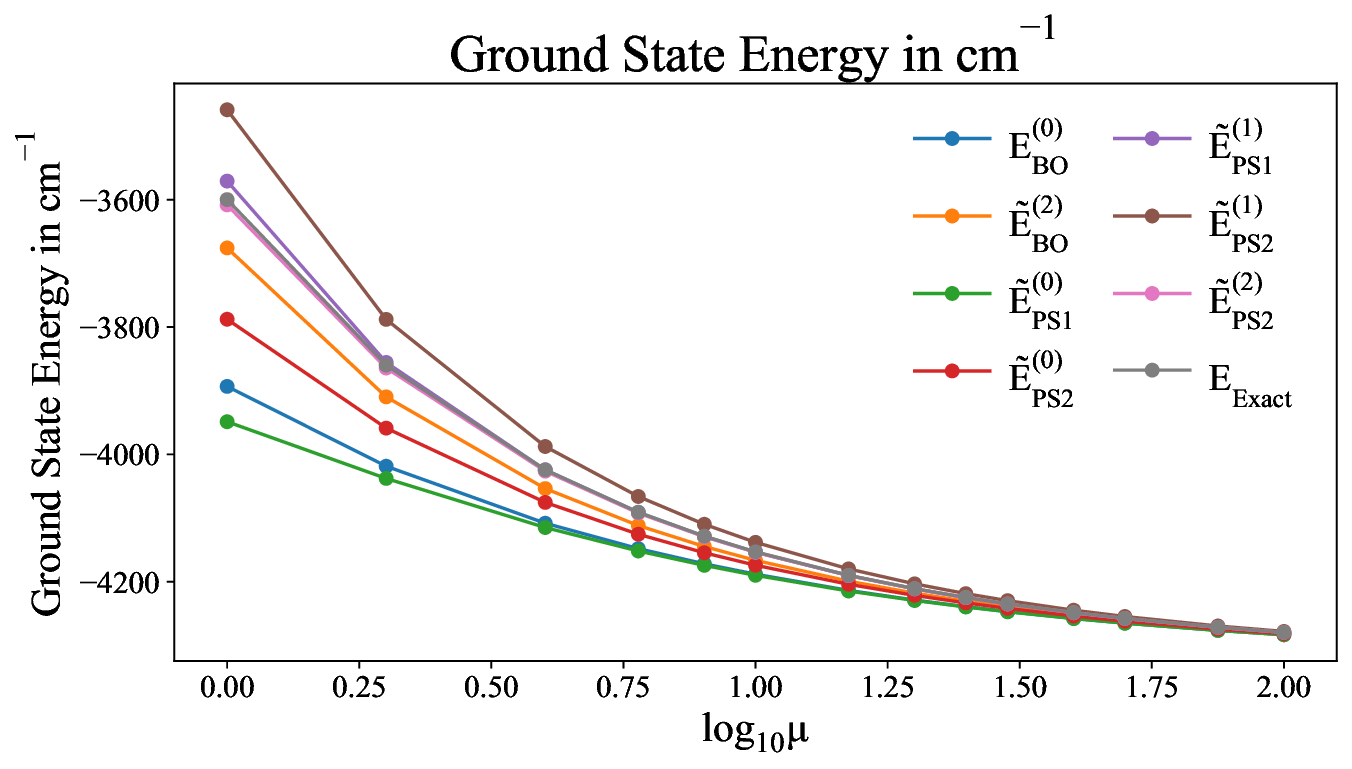}
\caption{\label{fig:abs_ge_womp} Ground state energy as a function of reduced mass $\mu$ without adding the mass polarization term. For this calculation, $\tilde{E}_{PS1}^{(1)}$ and $\tilde{E}_{PS2}^{(2)}$ yield the best answers.}
\end{figure}

\begin{figure}[H]
\includegraphics[width=0.9\linewidth, height=0.55\linewidth]{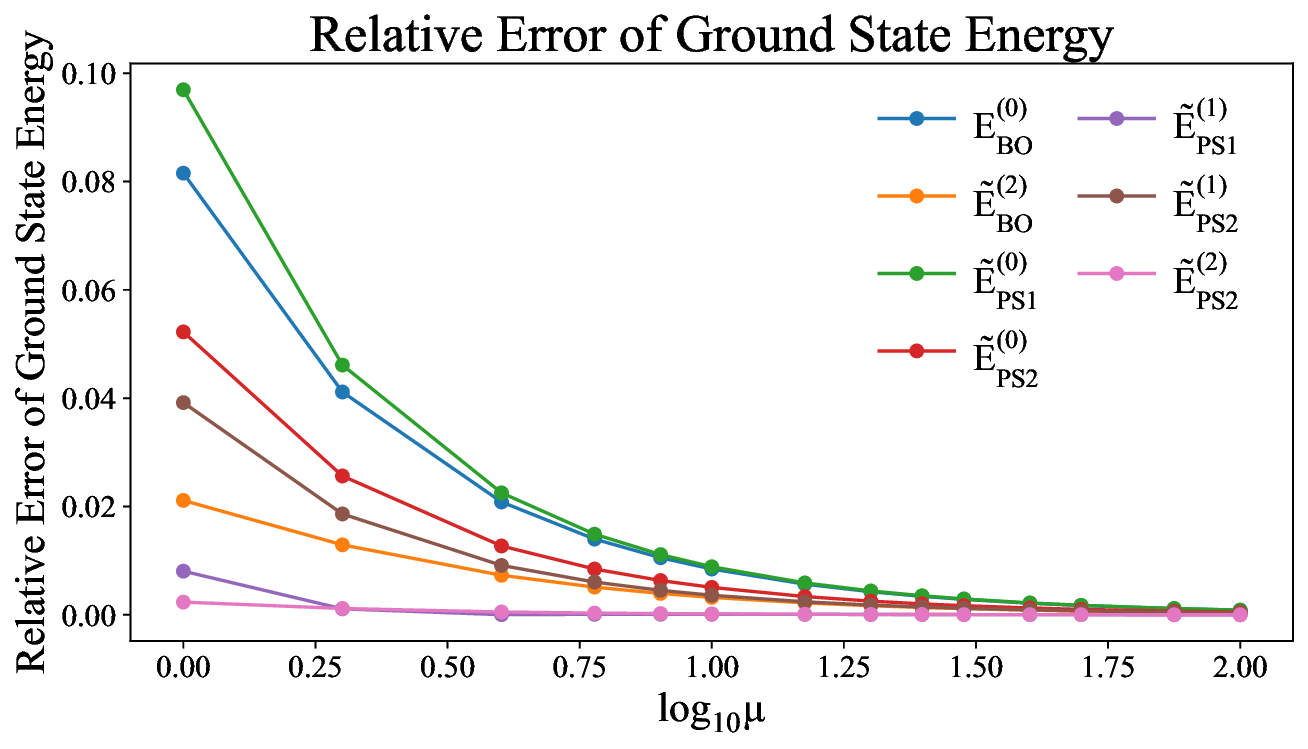}
\caption{\label{fig:re_ge_womp}  Relative error of the ground state energy as a function of reduced mass $\mu$ without adding the mass polarization term. We include this figure to make it visually easier for the reader to interpret Fig.  \ref{fig:abs_ge_womp} above.}
\end{figure}

\begin{figure}[H]
\includegraphics[width=0.9\linewidth, height=0.55\linewidth]{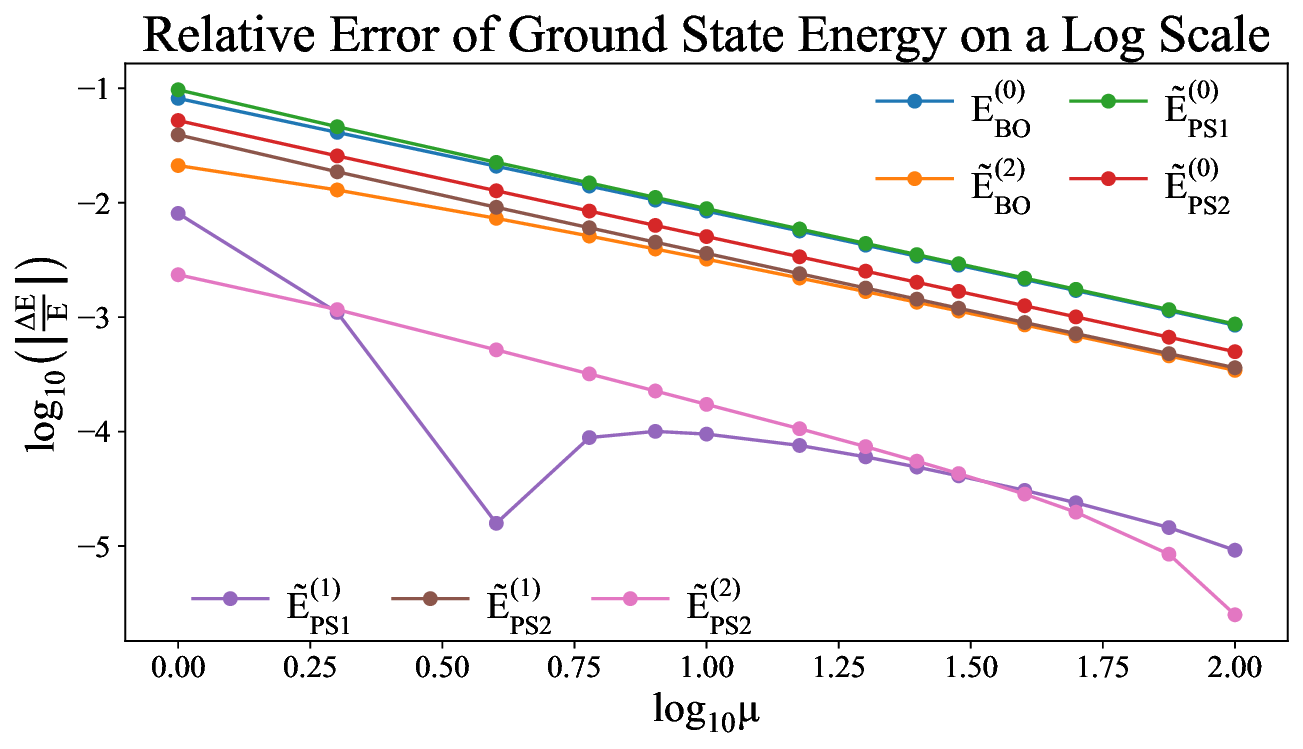}
\caption{\label{fig:logre_ge_womp} Relative error of the ground state energy as a function of reduced mass $\mu$ on a log scale without adding the mass polarization term. Same data as in Fig. \ref{fig:re_ge_womp} but replotted here for visual ease. Both $\tilde{E}_{PS2}^{(2)}$ and $\tilde{E}_{PS1}^{(1)}$ provide good estimates of the ground state energy when the mass polarization term is not included.}
\end{figure}

\begin{figure}[H]
\includegraphics[width=0.9\linewidth, height=0.55\linewidth]{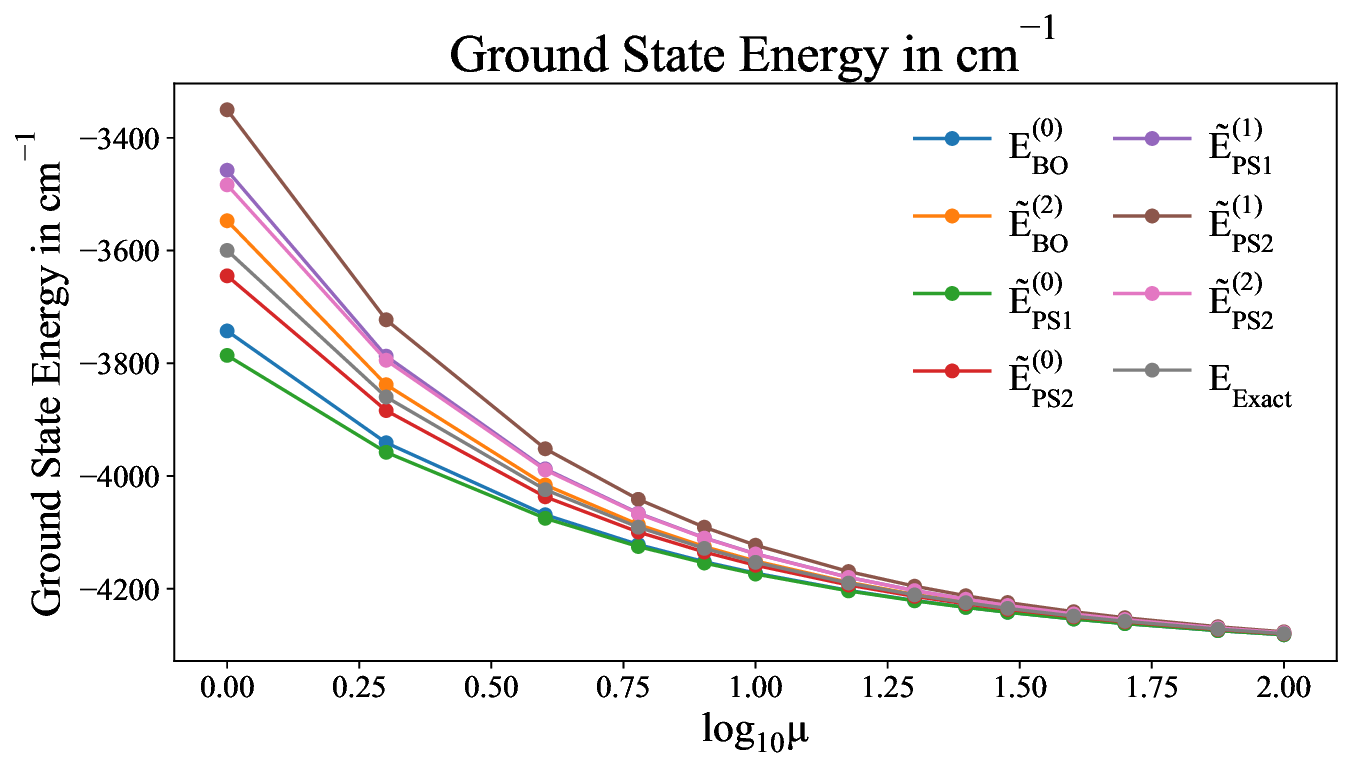}
\caption{\label{fig:abs_ge_wmp}Ground state energy as a function of reduced mass $\mu$ with the mass polarization term. }
\end{figure}

\begin{figure}[H]
\includegraphics[width=0.9\linewidth, height=0.55\linewidth]{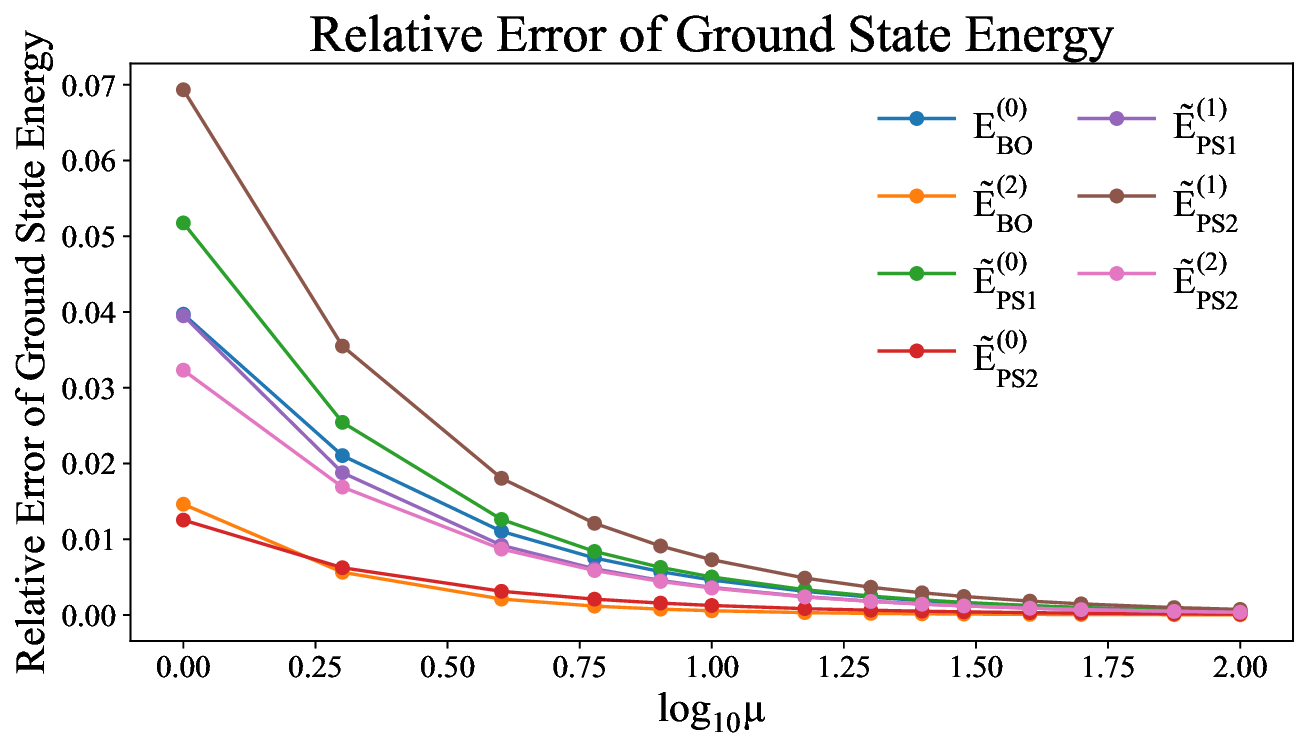}
\caption{\label{fig:re_ge_wmp} Relative error of the ground state energy as a function of reduced mass $\mu$ with the mass polarization term. With the mass polarization term, the best performance are given by $\tilde{E}_{PS2}^{(0)}$ and $\tilde{E}_{BO}^{(2)}$, which is quite different from what we found in Fig. \ref{fig:re_ge_womp}.}
\end{figure}

\begin{figure}[H]
\includegraphics[width=0.9\linewidth, height=0.55\linewidth]{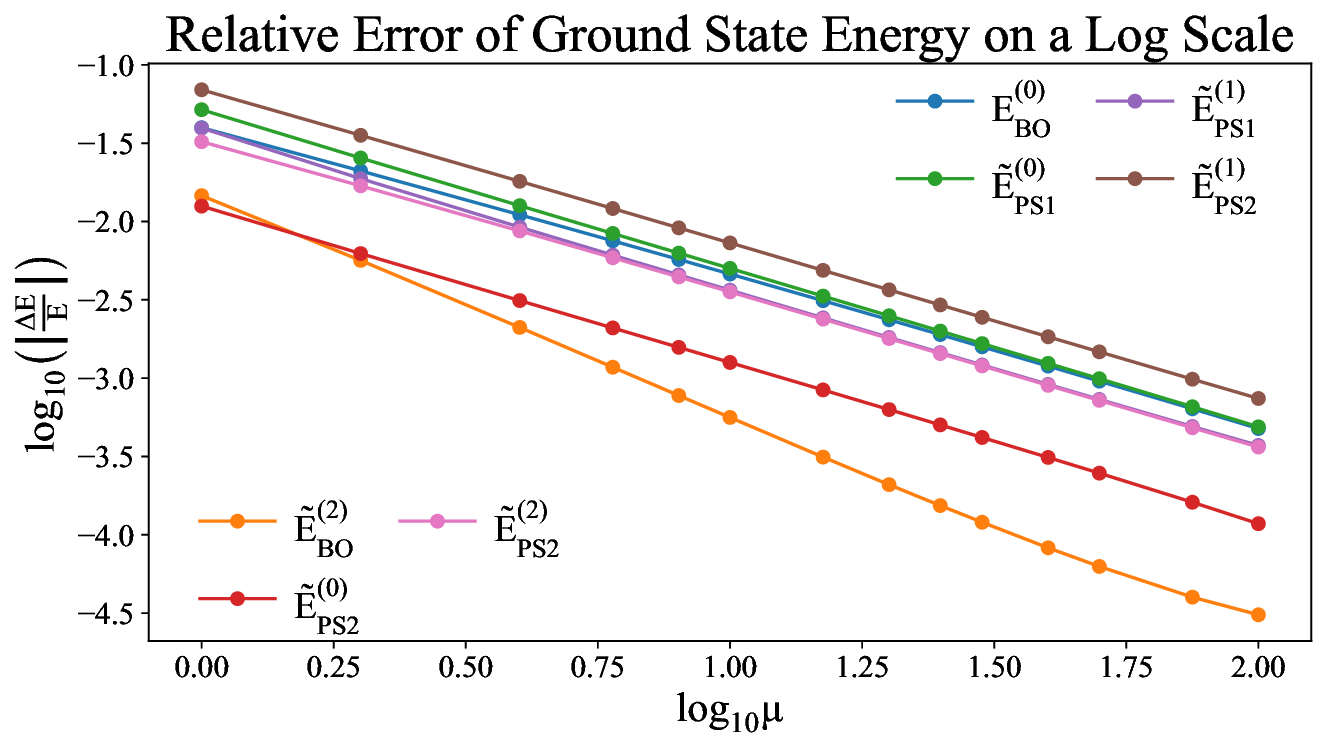}
\caption{\label{fig:logre_ge_wmp} Relative error of the ground state energy as a function of reduced mass $\mu$ on a log scale with the mass polarization term. Same data as Fig. \ref{fig:re_ge_wmp} but replotted here for visual ease.}
\end{figure}

To begin our analysis, in Figs. \ref{fig:abs_ge_womp}-\ref{fig:logre_ge_womp}, we plot the absolute ground state energy, the relative error in the ground state energy (on an absolute energy scale), and the relative error of the ground state energy (on logarithmic energy scale) when we do not include the mass polarization term; for Figs. \ref{fig:abs_ge_wmp}-\ref{fig:logre_ge_wmp}, we plot the same results where we do include the mass polarization term. As far as analyzing these results, one must be careful regarding the observable.  If one seeks the most accurate ground state energy, results $\tilde{E}_{PS1}^{(1)}$ and $\tilde{E}_{PS2}^{(2)}$ give the best answer without the mass polarization term. However, with the mass polarization term, $\tilde{E}_{PS2}^{(0)}$ and $\tilde{E}_{BO}^{(2)}$ seem to give the best answer.

As far as scaling is concerned, note that the $\tilde{E}_{BO}^{(2)}$ goes to zero as $\mu^{-3/2}$ in Fig. \ref{fig:logre_ge_wmp}; for this method, we include all anharmonicity and mass polarization, while the inaccuracy is dominated by the $K_{22}$ term (which arises from the second order correction in Eq. \ref{eq:hlf:expansion}); for more details, see Ref. \citenum{littlejohn:2024:jcp:moyal}. Interestingly, for this set of data and for Figs. \ref{fig:abs_ge_wmp}-\ref{fig:logre_ge_wmp}, the scaling of the PS data never scales higher than $\mu^{-1}$, likely because our second order correction  (in Eq. \ref{eq:naive2}) is only approximate.

\subsection{Vibrational Energies are much less affected}

\begin{figure}[H]
\includegraphics[width=0.9\linewidth, height=0.55\linewidth]{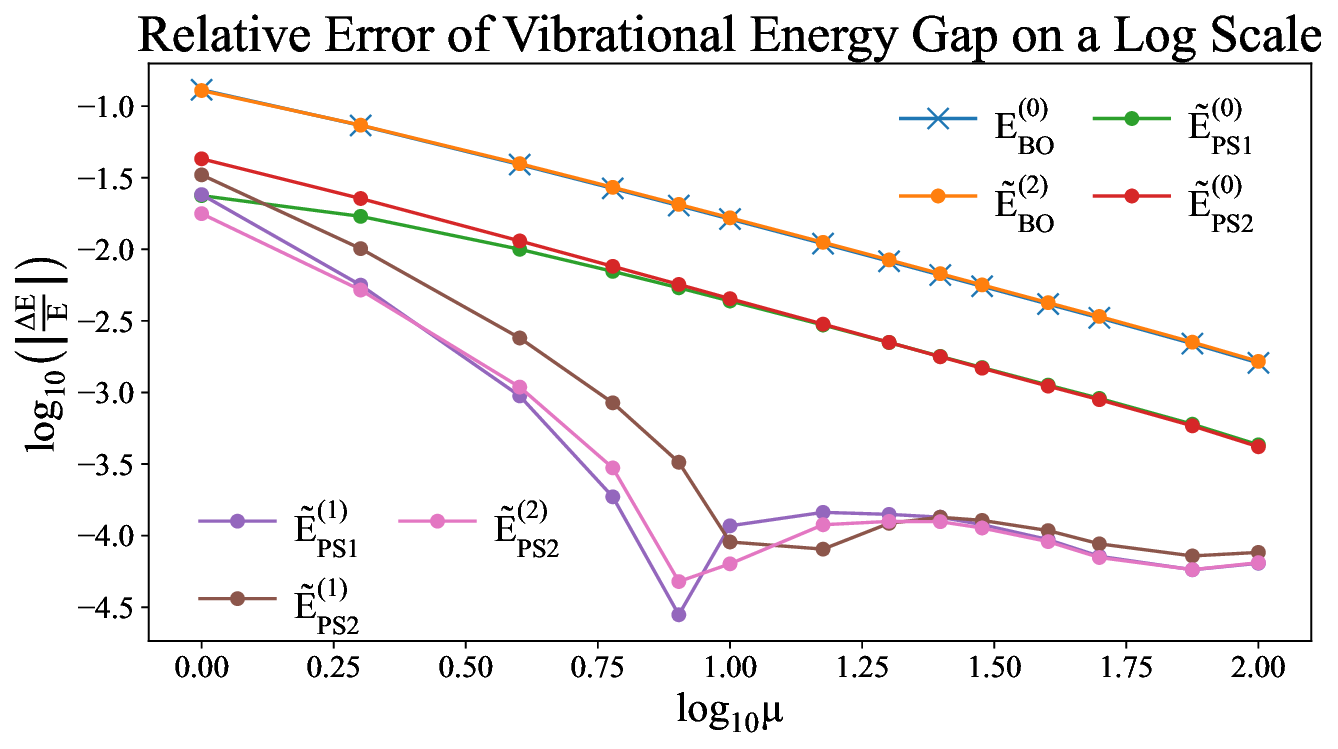}
\caption{\label{fig:logre_vib_wmp}  Relative error of vibrational energy gap as a function of reduced mass $\mu$ on a log scale with the mass polarization term.  Note that $\tilde{E}_{PS2}^{(2)}$ and $\tilde{E}_{PS1}^{(1)}$ are still the best performing, which is very similar to Fig.\ref{fig:logre_vib_womp}.}
\end{figure}

As reported in Fig. \ref{fig:abs_ge_womp}-\ref{fig:logre_ge_wmp}, the absolute energy scales of the ground state energy can depend sensitively on the moving frame, and the method of choice. Luckily, vibrational energies are more important than absolute energies, and we note that
vibrational energies are much less affected by the mass polarization term,  and $\tilde{E}_{PS1}^{(1)}$, $\tilde{E}_{PS2}^{(2)}$ and $\tilde{E}_{PS2}^{(1)}$ always give the best answers. 
To prove this point, in Fig. \ref{fig:logre_vib_wmp}, we plot the relative error on a log scale of the  vibrational energies when we include the mass polarization term. Indeed, these results are very similar to the results without the mass polarization term (in Fig. \ref{fig:abs_vib_womp}-\ref{fig:logre_vib_womp}).


 \section {Conclusions}
\label{sec:conclusion}
We have developed a formal but practical theory of phase space electronic Hamiltonians, whereby at zeroth order one diagonalizes an electronic operator parameterized by both $R$ and $P$, and thereafter one can continue to improve the resulting eigenvalues perturbatively approaching the exact answer.   Unlike the standard BO representation, where the expansion of the total Hamiltonian terminates at second order in $\hbar$, with a phase space formalism, the exact Hamiltonian must be written down as a non-terminating perturbative expansion  in $\hbar$ (as in Eq. \ref{eq:PS:expand_in_hbar}) following Littlejohn-Flynn theory.
At the end of the day,  although the formal theory in Sec. \ref{sec:ps} above does not make any statement regarding how to choose a $\hGamma$ operator, our analysis does show how to mathematically ground the hypothesized phase space electronic Hamiltonians from Ref. \citenum{tao2025basis}  within a rigorous framework.  Even without a unique $\hGamma$, phase space electronic structure theory is not an {\em ad hoc} approach. 

Turning to our results,  in Sec. \ref{sec:model}, we have shown that a first order correction to phase space theory can strongly improve vibrational energies -- as long as we go beyond the harmonic limit, relative energies can improve by a factor of 10 when we go to first order in $\hbar$, though at the cost of taking derivatives with respect to $R$ and $P$. As a practical matter, we are unsure whether this first-order correction will be of interest for larger systems given the need for such derivatives. 
Another important question surrounds the expansion in Eq. \ref{eq:PS:expand_in_hbar}. Note that Eq. \ref{eq:PS:expand_in_hbar} is perturbative, as can be ascertained from the denominator in Eq. \ref{eqn:L_w1ij}. Thus, Eq. \ref{eq:PS:expand_in_hbar} is only expected to be accurate in the limit that the ground electronic state is reasonably well separated from the higher electronic states.  Looking  forward, it will also be crucial in the future to figure out how to rigorously work with non-perturbatively coupled, nearly degenerate systems (e.g. spin systems and systems with multiconfigurational character), where one needs to avoid the denominators in Eq. \ref{eqn:L_w1ij}.

In the end, though a great deal of future work remains to be done, 
the present paper shows how to recover exact quantum eigenvalues starting with phase space electronic states. As such, this paper represents an important numerical validation of the entire phase space approach and is a strong endorsement of developing non-Born Oppenheimer electronic structure theory.

\bibliography{ref}

\providecommand{\latin}[1]{#1}
\makeatletter
\providecommand{\doi}
  {\begingroup\let\do\@makeother\dospecials
  \catcode`\{=1 \catcode`\}=2 \doi@aux}
\providecommand{\doi@aux}[1]{\endgroup\texttt{#1}}
\makeatother
\providecommand*\mcitethebibliography{\thebibliography}
\csname @ifundefined\endcsname{endmcitethebibliography}  {\let\endmcitethebibliography\endthebibliography}{}
\begin{mcitethebibliography}{46}
\providecommand*\natexlab[1]{#1}
\providecommand*\mciteSetBstSublistMode[1]{}
\providecommand*\mciteSetBstMaxWidthForm[2]{}
\providecommand*\mciteBstWouldAddEndPuncttrue
  {\def\EndOfBibitem{\unskip.}}
\providecommand*\mciteBstWouldAddEndPunctfalse
  {\let\EndOfBibitem\relax}
\providecommand*\mciteSetBstMidEndSepPunct[3]{}
\providecommand*\mciteSetBstSublistLabelBeginEnd[3]{}
\providecommand*\EndOfBibitem{}
\mciteSetBstSublistMode{f}
\mciteSetBstMaxWidthForm{subitem}{(\alph{mcitesubitemcount})}
\mciteSetBstSublistLabelBeginEnd
  {\mcitemaxwidthsubitemform\space}
  {\relax}
  {\relax}

\bibitem[Colthup(2012)]{colthup2012introduction}
Colthup,~N. \emph{Introduction to infrared and Raman spectroscopy}; Elsevier, 2012\relax
\mciteBstWouldAddEndPuncttrue
\mciteSetBstMidEndSepPunct{\mcitedefaultmidpunct}
{\mcitedefaultendpunct}{\mcitedefaultseppunct}\relax
\EndOfBibitem
\bibitem[Brandt \latin{et~al.}(2016)Brandt, Keller, and Frontiera]{frontiera:2016:jpcl:raman_hot}
Brandt,~N.~C.; Keller,~E.~L.; Frontiera,~R.~R. Ultrafast Surface-Enhanced Raman Probing of the Role of Hot Electrons in Plasmon-Driven Chemistry. \emph{Journal of Physical Chemistry Letters} \textbf{2016}, \emph{7}, 3179--3185\relax
\mciteBstWouldAddEndPuncttrue
\mciteSetBstMidEndSepPunct{\mcitedefaultmidpunct}
{\mcitedefaultendpunct}{\mcitedefaultseppunct}\relax
\EndOfBibitem
\bibitem[Heitler and London(1927)Heitler, and London]{heitler1927wechselwirkung}
Heitler,~W.; London,~F. Wechselwirkung neutraler Atome und hom{\"o}opolare Bindung nach der Quantenmechanik. \emph{Zeitschrift f{\"u}r Physik} \textbf{1927}, \emph{44}, 455--472\relax
\mciteBstWouldAddEndPuncttrue
\mciteSetBstMidEndSepPunct{\mcitedefaultmidpunct}
{\mcitedefaultendpunct}{\mcitedefaultseppunct}\relax
\EndOfBibitem
\bibitem[Born and Huang(1996)Born, and Huang]{born1996dynamical}
Born,~M.; Huang,~K. \emph{Dynamical theory of crystal lattices}; Oxford university press, 1996\relax
\mciteBstWouldAddEndPuncttrue
\mciteSetBstMidEndSepPunct{\mcitedefaultmidpunct}
{\mcitedefaultendpunct}{\mcitedefaultseppunct}\relax
\EndOfBibitem
\bibitem[Cederbaum(2004)]{cederbaum:review:conicalbook}
Cederbaum,~L.~S. In \emph{Conical Intersections: Electronic Structure, Dynamics and Spectroscopy}; Domcke,~W., Yarkony,~D.~R., Koppel,~H., Eds.; World Scientific Publishing Co.: New Jersey, 2004; pp 3--40\relax
\mciteBstWouldAddEndPuncttrue
\mciteSetBstMidEndSepPunct{\mcitedefaultmidpunct}
{\mcitedefaultendpunct}{\mcitedefaultseppunct}\relax
\EndOfBibitem
\bibitem[Bernath(2020)]{bernath2020spectra}
Bernath,~P.~F. \emph{Spectra of atoms and molecules}; Oxford university press, 2020\relax
\mciteBstWouldAddEndPuncttrue
\mciteSetBstMidEndSepPunct{\mcitedefaultmidpunct}
{\mcitedefaultendpunct}{\mcitedefaultseppunct}\relax
\EndOfBibitem
\bibitem[Lefebvre-Brion and Field(2004)Lefebvre-Brion, and Field]{lefebvre2004spectra}
Lefebvre-Brion,~H.; Field,~R.~W. \emph{The spectra and dynamics of diatomic molecules: revised and enlarged edition}; Elsevier, 2004\relax
\mciteBstWouldAddEndPuncttrue
\mciteSetBstMidEndSepPunct{\mcitedefaultmidpunct}
{\mcitedefaultendpunct}{\mcitedefaultseppunct}\relax
\EndOfBibitem
\bibitem[Bunker and Moss(1980)Bunker, and Moss]{bunker1980effect}
Bunker,~P.; Moss,~R. The effect of the breakdown of the Born-Oppenheimer approximation on the rotation-vibration Hamiltonian of a triatomic molecule. \emph{Journal of Molecular Spectroscopy} \textbf{1980}, \emph{80}, 217--228\relax
\mciteBstWouldAddEndPuncttrue
\mciteSetBstMidEndSepPunct{\mcitedefaultmidpunct}
{\mcitedefaultendpunct}{\mcitedefaultseppunct}\relax
\EndOfBibitem
\bibitem[Bunker and Jensen(2006)Bunker, and Jensen]{bunker2006molecular}
Bunker,~P.~R.; Jensen,~P. \emph{Molecular symmetry and spectroscopy}; NRC research press, 2006; Vol. 46853\relax
\mciteBstWouldAddEndPuncttrue
\mciteSetBstMidEndSepPunct{\mcitedefaultmidpunct}
{\mcitedefaultendpunct}{\mcitedefaultseppunct}\relax
\EndOfBibitem
\bibitem[Schwenke(2001)]{schwenke2001beyond}
Schwenke,~D.~W. Beyond the Potential Energy Surface: Ab initio Corrections to the Born- Oppenheimer Approximation for H2O. \emph{The Journal of Physical Chemistry A} \textbf{2001}, \emph{105}, 2352--2360\relax
\mciteBstWouldAddEndPuncttrue
\mciteSetBstMidEndSepPunct{\mcitedefaultmidpunct}
{\mcitedefaultendpunct}{\mcitedefaultseppunct}\relax
\EndOfBibitem
\bibitem[Scherrer \latin{et~al.}(2017)Scherrer, Agostini, Sebastiani, Gross, and Vuilleumier]{scherrer2017mass}
Scherrer,~A.; Agostini,~F.; Sebastiani,~D.; Gross,~E.; Vuilleumier,~R. On the mass of atoms in molecules: Beyond the Born-Oppenheimer approximation. \emph{Physical Review X} \textbf{2017}, \emph{7}, 031035\relax
\mciteBstWouldAddEndPuncttrue
\mciteSetBstMidEndSepPunct{\mcitedefaultmidpunct}
{\mcitedefaultendpunct}{\mcitedefaultseppunct}\relax
\EndOfBibitem
\bibitem[Baer(1975)]{mbaer:1975:cpl}
Baer,~M. Adiabatic and diabatic representations for atom-molecule collisions: Treatment of the collinear arrangement. \emph{Chemical Physics Letters} \textbf{1975}, \emph{35}, 112\relax
\mciteBstWouldAddEndPuncttrue
\mciteSetBstMidEndSepPunct{\mcitedefaultmidpunct}
{\mcitedefaultendpunct}{\mcitedefaultseppunct}\relax
\EndOfBibitem
\bibitem[Mead and Truhlar(1982)Mead, and Truhlar]{meadtruhlar:1982:conditions_diabatic}
Mead,~C.~A.; Truhlar,~D.~G. Conditions for the definition of a strictly diabatic electronic basis for molecular systems. \emph{Journal of Chemical Physics} \textbf{1982}, \emph{77}, 6090--6098\relax
\mciteBstWouldAddEndPuncttrue
\mciteSetBstMidEndSepPunct{\mcitedefaultmidpunct}
{\mcitedefaultendpunct}{\mcitedefaultseppunct}\relax
\EndOfBibitem
\bibitem[Bubin \latin{et~al.}(2013)Bubin, Pavanello, Tung, Sharkey, and Adamowicz]{bubin2013born}
Bubin,~S.; Pavanello,~M.; Tung,~W.-C.; Sharkey,~K.~L.; Adamowicz,~L. Born--Oppenheimer and non-Born--Oppenheimer, atomic and molecular calculations with explicitly correlated Gaussians. \emph{Chemical reviews} \textbf{2013}, \emph{113}, 36--79\relax
\mciteBstWouldAddEndPuncttrue
\mciteSetBstMidEndSepPunct{\mcitedefaultmidpunct}
{\mcitedefaultendpunct}{\mcitedefaultseppunct}\relax
\EndOfBibitem
\bibitem[Tao \latin{et~al.}(2025)Tao, Qiu, Bian, Duston, Bradbury, and Subotnik]{tao2025basis}
Tao,~Z.; Qiu,~T.; Bian,~X.; Duston,~T.; Bradbury,~N.; Subotnik,~J.~E. A Basis-free phase space electronic Hamiltonian that recovers beyond Born--Oppenheimer electronic momentum and current density. \emph{The Journal of Chemical Physics} \textbf{2025}, \emph{162}\relax
\mciteBstWouldAddEndPuncttrue
\mciteSetBstMidEndSepPunct{\mcitedefaultmidpunct}
{\mcitedefaultendpunct}{\mcitedefaultseppunct}\relax
\EndOfBibitem
\bibitem[Bian \latin{et~al.}(0)Bian, Khan, Duston, Rawlinson, Littlejohn, and Subotnik]{bian:2025:jctc:wigner_vibrations}
Bian,~X.; Khan,~C.; Duston,~T.; Rawlinson,~J.; Littlejohn,~R.~G.; Subotnik,~J.~E. A Phase-Space View of Vibrational Energies without the Born–Oppenheimer Framework. \emph{Journal of Chemical Theory and Computation} \textbf{0}, \emph{0}, null, PMID: 40072941\relax
\mciteBstWouldAddEndPuncttrue
\mciteSetBstMidEndSepPunct{\mcitedefaultmidpunct}
{\mcitedefaultendpunct}{\mcitedefaultseppunct}\relax
\EndOfBibitem
\bibitem[Tao \latin{et~al.}(2024)Tao, Qiu, Bhati, Bian, Duston, Rawlinson, Littlejohn, and Subotnik]{tao2024practical}
Tao,~Z.; Qiu,~T.; Bhati,~M.; Bian,~X.; Duston,~T.; Rawlinson,~J.; Littlejohn,~R.~G.; Subotnik,~J.~E. Practical phase-space electronic Hamiltonians for ab initio dynamics. \emph{The Journal of Chemical Physics} \textbf{2024}, \emph{160}\relax
\mciteBstWouldAddEndPuncttrue
\mciteSetBstMidEndSepPunct{\mcitedefaultmidpunct}
{\mcitedefaultendpunct}{\mcitedefaultseppunct}\relax
\EndOfBibitem
\bibitem[Nafie(1983)]{nafie1983adiabatic}
Nafie,~L.~A. Adiabatic molecular properties beyond the Born--Oppenheimer approximation. Complete adiabatic wave functions and vibrationally induced electronic current density. \emph{The Journal of chemical physics} \textbf{1983}, \emph{79}, 4950--4957\relax
\mciteBstWouldAddEndPuncttrue
\mciteSetBstMidEndSepPunct{\mcitedefaultmidpunct}
{\mcitedefaultendpunct}{\mcitedefaultseppunct}\relax
\EndOfBibitem
\bibitem[Patchkovskii(2012)]{patchkovskii:2012:jcp:electronic_current}
Patchkovskii,~S. Electronic currents and Born-Oppenheimer molecular dynamics. \emph{Journal of Chemical Physics} \textbf{2012}, \emph{137}, 084109\relax
\mciteBstWouldAddEndPuncttrue
\mciteSetBstMidEndSepPunct{\mcitedefaultmidpunct}
{\mcitedefaultendpunct}{\mcitedefaultseppunct}\relax
\EndOfBibitem
\bibitem[Hanasaki and Takatsuka(2021)Hanasaki, and Takatsuka]{takatsuka:2021:jcp:flux_conservation}
Hanasaki,~K.; Takatsuka,~K. On the molecular electronic flux: Role of nonadiabaticity and violation of conservation. \emph{Journal of Chemical Physics} \textbf{2021}, \emph{154}\relax
\mciteBstWouldAddEndPuncttrue
\mciteSetBstMidEndSepPunct{\mcitedefaultmidpunct}
{\mcitedefaultendpunct}{\mcitedefaultseppunct}\relax
\EndOfBibitem
\bibitem[Qiu \latin{et~al.}(2024)Qiu, Bhati, Tao, Bian, Rawlinson, Littlejohn, and Subotnik]{tian:2024:jcp:erf}
Qiu,~T.; Bhati,~M.; Tao,~Z.; Bian,~X.; Rawlinson,~J.; Littlejohn,~R.~G.; Subotnik,~J.~E. A Simple One-Electron Expression for Electron Rotational Factors. 2024; J. Chem. Phys., in press\relax
\mciteBstWouldAddEndPuncttrue
\mciteSetBstMidEndSepPunct{\mcitedefaultmidpunct}
{\mcitedefaultendpunct}{\mcitedefaultseppunct}\relax
\EndOfBibitem
\bibitem[Littlejohn \latin{et~al.}(2023)Littlejohn, Rawlinson, and Subotnik]{littlejohn:2023:jcp:angmom}
Littlejohn,~R.; Rawlinson,~J.; Subotnik,~J. Representation and conservation of angular momentum in the Born--Oppenheimer theory of polyatomic molecules. \emph{Journal of Chemical Physics} \textbf{2023}, \emph{158}, 104302\relax
\mciteBstWouldAddEndPuncttrue
\mciteSetBstMidEndSepPunct{\mcitedefaultmidpunct}
{\mcitedefaultendpunct}{\mcitedefaultseppunct}\relax
\EndOfBibitem
\bibitem[Bian \latin{et~al.}(2023)Bian, Tao, Wu, Rawlinson, Littlejohn, and Subotnik]{bian2023total}
Bian,~X.; Tao,~Z.; Wu,~Y.; Rawlinson,~J.; Littlejohn,~R.~G.; Subotnik,~J.~E. Total angular momentum conservation in ab initio Born-Oppenheimer molecular dynamics. \emph{Physical Review B} \textbf{2023}, \emph{108}, L220304\relax
\mciteBstWouldAddEndPuncttrue
\mciteSetBstMidEndSepPunct{\mcitedefaultmidpunct}
{\mcitedefaultendpunct}{\mcitedefaultseppunct}\relax
\EndOfBibitem
\bibitem[Duston \latin{et~al.}(2024)Duston, Tao, Bian, Bhati, Rawlinson, Littlejohn, Pei, Shao, and Subotnik]{duston2024phase}
Duston,~T.; Tao,~Z.; Bian,~X.; Bhati,~M.; Rawlinson,~J.; Littlejohn,~R.~G.; Pei,~Z.; Shao,~Y.; Subotnik,~J.~E. A phase-space electronic Hamiltonian for vibrational circular dichroism. \emph{Journal of Chemical Theory and Computation} \textbf{2024}, \emph{20}, 7904--7921\relax
\mciteBstWouldAddEndPuncttrue
\mciteSetBstMidEndSepPunct{\mcitedefaultmidpunct}
{\mcitedefaultendpunct}{\mcitedefaultseppunct}\relax
\EndOfBibitem
\bibitem[Tao \latin{et~al.}(2024)Tao, Duston, Pei, Shao, Rawlinson, Littlejohn, and Subotnik]{tao2024electronic}
Tao,~Z.; Duston,~T.; Pei,~Z.; Shao,~Y.; Rawlinson,~J.; Littlejohn,~R.; Subotnik,~J.~E. An electronic phase-space Hamiltonian approach for electronic current density and vibrational circular dichroism. \emph{The Journal of Chemical Physics} \textbf{2024}, \emph{161}\relax
\mciteBstWouldAddEndPuncttrue
\mciteSetBstMidEndSepPunct{\mcitedefaultmidpunct}
{\mcitedefaultendpunct}{\mcitedefaultseppunct}\relax
\EndOfBibitem
\bibitem[Stephens(1985)]{stephens:1985:jpcc_vcd}
Stephens,~P.~J. Theory of vibrational circular dichroism. \emph{Journal of Physical Chemistry} \textbf{1985}, \emph{89}, 748--752\relax
\mciteBstWouldAddEndPuncttrue
\mciteSetBstMidEndSepPunct{\mcitedefaultmidpunct}
{\mcitedefaultendpunct}{\mcitedefaultseppunct}\relax
\EndOfBibitem
\bibitem[D’Alessio and Polkovnikov(2014)D’Alessio, and Polkovnikov]{d2014emergent}
D’Alessio,~L.; Polkovnikov,~A. Emergent Newtonian dynamics and the geometric origin of mass. \emph{Annals of Physics} \textbf{2014}, \emph{345}, 141--165\relax
\mciteBstWouldAddEndPuncttrue
\mciteSetBstMidEndSepPunct{\mcitedefaultmidpunct}
{\mcitedefaultendpunct}{\mcitedefaultseppunct}\relax
\EndOfBibitem
\bibitem[Barrera \latin{et~al.}()Barrera, Arovas, Chandran, and Polkovnikov]{polkovnikov:2025:moving_boa}
Barrera,~B.; Arovas,~D.; Chandran,~A.; Polkovnikov,~A. The Moving Born-Oppenheimer Approximation. arXiv:2502.17557\relax
\mciteBstWouldAddEndPuncttrue
\mciteSetBstMidEndSepPunct{\mcitedefaultmidpunct}
{\mcitedefaultendpunct}{\mcitedefaultseppunct}\relax
\EndOfBibitem
\bibitem[Blount(1962)]{blount1962}
Blount,~E. Bloch electrons in a magnetic field. \emph{Physical Review} \textbf{1962}, \emph{126}, 1636\relax
\mciteBstWouldAddEndPuncttrue
\mciteSetBstMidEndSepPunct{\mcitedefaultmidpunct}
{\mcitedefaultendpunct}{\mcitedefaultseppunct}\relax
\EndOfBibitem
\bibitem[Littlejohn and Flynn(1991)Littlejohn, and Flynn]{littlejohn1991}
Littlejohn,~R.~G.; Flynn,~W.~G. Geometric phases and the Bohr-Sommerfeld quantization of multicomponent wave fields. \emph{Physical review letters} \textbf{1991}, \emph{66}, 2839\relax
\mciteBstWouldAddEndPuncttrue
\mciteSetBstMidEndSepPunct{\mcitedefaultmidpunct}
{\mcitedefaultendpunct}{\mcitedefaultseppunct}\relax
\EndOfBibitem
\bibitem[Littlejohn and Flynn(1991)Littlejohn, and Flynn]{littlejohn:flynn:1991:pra:coriolis}
Littlejohn,~R.~G.; Flynn,~W.~G. Geometric phases in the asymptotic theory of coupled wave equations. \emph{Phys. Rev. A} \textbf{1991}, \emph{44}, 5239--5256\relax
\mciteBstWouldAddEndPuncttrue
\mciteSetBstMidEndSepPunct{\mcitedefaultmidpunct}
{\mcitedefaultendpunct}{\mcitedefaultseppunct}\relax
\EndOfBibitem
\bibitem[Littlejohn \latin{et~al.}(2024)Littlejohn, Rawlinson, and Subotnik]{littlejohn:2024:jcp:moyal}
Littlejohn,~R.; Rawlinson,~J.; Subotnik,~J. {Diagonalizing the Born–Oppenheimer Hamiltonian via Moyal perturbation theory, nonadiabatic corrections, and translational degrees of freedom}. \emph{Journal of Chemical Physics} \textbf{2024}, \emph{160}, 114103\relax
\mciteBstWouldAddEndPuncttrue
\mciteSetBstMidEndSepPunct{\mcitedefaultmidpunct}
{\mcitedefaultendpunct}{\mcitedefaultseppunct}\relax
\EndOfBibitem
\bibitem[Teufel(2003)]{teufel2003}
Teufel,~S. \emph{Adiabatic perturbation theory in quantum dynamics}; Springer Science \& Business Media, 2003\relax
\mciteBstWouldAddEndPuncttrue
\mciteSetBstMidEndSepPunct{\mcitedefaultmidpunct}
{\mcitedefaultendpunct}{\mcitedefaultseppunct}\relax
\EndOfBibitem
\bibitem[M{\'a}tyus and Teufel(2019)M{\'a}tyus, and Teufel]{matyus2019}
M{\'a}tyus,~E.; Teufel,~S. Effective non-adiabatic Hamiltonians for the quantum nuclear motion over coupled electronic states. \emph{The Journal of Chemical Physics} \textbf{2019}, \emph{151}\relax
\mciteBstWouldAddEndPuncttrue
\mciteSetBstMidEndSepPunct{\mcitedefaultmidpunct}
{\mcitedefaultendpunct}{\mcitedefaultseppunct}\relax
\EndOfBibitem
\bibitem[Martens and Fang(1997)Martens, and Fang]{martens:1997:partwig}
Martens,~C.~C.; Fang,~J.~Y. Semiclassical-limit molecular dynamics on multiple electronic surfaces. \emph{Journal of Chemical Physics} \textbf{1997}, \emph{106}, 4918--4930\relax
\mciteBstWouldAddEndPuncttrue
\mciteSetBstMidEndSepPunct{\mcitedefaultmidpunct}
{\mcitedefaultendpunct}{\mcitedefaultseppunct}\relax
\EndOfBibitem
\bibitem[Donoso and Martens(1998)Donoso, and Martens]{martens:1998:partwig}
Donoso,~A.; Martens,~C.~C. Simulation of coherent nonadiabatic dynamics using classical trajectories. \emph{Journal of Physical Chemistry A} \textbf{1998}, \emph{102}, 4291--4300\relax
\mciteBstWouldAddEndPuncttrue
\mciteSetBstMidEndSepPunct{\mcitedefaultmidpunct}
{\mcitedefaultendpunct}{\mcitedefaultseppunct}\relax
\EndOfBibitem
\bibitem[Kapral and Ciccotti(1999)Kapral, and Ciccotti]{kapral:1999:jcp}
Kapral,~R.; Ciccotti,~G. Mixed quantum-classical dynamics. \emph{Journal of Chemical Physics} \textbf{1999}, \emph{110}, 8919--8929\relax
\mciteBstWouldAddEndPuncttrue
\mciteSetBstMidEndSepPunct{\mcitedefaultmidpunct}
{\mcitedefaultendpunct}{\mcitedefaultseppunct}\relax
\EndOfBibitem
\bibitem[Grunwald \latin{et~al.}(2009)Grunwald, Kelly, and Kapral]{kapral:2009:burghardt}
Grunwald,~R.; Kelly,~A.; Kapral,~R. In \emph{Energy Transfer Dynamics in Biomaterial Systems}; Burghardt,~I., Ed.; Springer-Verlag: Berlin, 2009; p 383\relax
\mciteBstWouldAddEndPuncttrue
\mciteSetBstMidEndSepPunct{\mcitedefaultmidpunct}
{\mcitedefaultendpunct}{\mcitedefaultseppunct}\relax
\EndOfBibitem
\bibitem[Kelly and Kapral(2010)Kelly, and Kapral]{kelly:2010:jcp}
Kelly,~A.; Kapral,~R. Quantum-classical description of environmental effects on electronic dynamics at conical intersections. \emph{Journal of Chemical Physics} \textbf{2010}, \emph{133}, 084502\relax
\mciteBstWouldAddEndPuncttrue
\mciteSetBstMidEndSepPunct{\mcitedefaultmidpunct}
{\mcitedefaultendpunct}{\mcitedefaultseppunct}\relax
\EndOfBibitem
\bibitem[Nielsen \latin{et~al.}(2000)Nielsen, Kapral, and Ciccotti]{kapral:2000:jcp}
Nielsen,~S.; Kapral,~R.; Ciccotti,~G. Mixed quantum-classical surface hopping dynamics. \emph{Journal of Chemical Physics} \textbf{2000}, \emph{112}, 6543--6553\relax
\mciteBstWouldAddEndPuncttrue
\mciteSetBstMidEndSepPunct{\mcitedefaultmidpunct}
{\mcitedefaultendpunct}{\mcitedefaultseppunct}\relax
\EndOfBibitem
\bibitem[Huo and Coker(2012)Huo, and Coker]{coker:2012:iterative}
Huo,~P.; Coker,~D. Consistent schemes for non-adiabatic dynamics derived from partial linearized density matrix propagation. \emph{Journal of Chemical Physics} \textbf{2012}, \emph{137}, 22A535\relax
\mciteBstWouldAddEndPuncttrue
\mciteSetBstMidEndSepPunct{\mcitedefaultmidpunct}
{\mcitedefaultendpunct}{\mcitedefaultseppunct}\relax
\EndOfBibitem
\bibitem[Shi and Geva(2003)Shi, and Geva]{geva:shi:2003:semiclassical_relaxation:jpca}
Shi,~Q.; Geva,~E. Semiclassical Theory of Vibrational Energy Relaxation in the Condensed Phase. \emph{Journal of Physical Chemistry A} \textbf{2003}, \emph{107}, 9059--9069\relax
\mciteBstWouldAddEndPuncttrue
\mciteSetBstMidEndSepPunct{\mcitedefaultmidpunct}
{\mcitedefaultendpunct}{\mcitedefaultseppunct}\relax
\EndOfBibitem
\bibitem[Marinica \latin{et~al.}(2006)Marinica, Gaigeot, and Borgis]{marinica2006generating}
Marinica,~D.~C.; Gaigeot,~M.-P.; Borgis,~D. Generating approximate Wigner distributions using Gaussian phase packets propagation in imaginary time. \emph{Chemical physics letters} \textbf{2006}, \emph{423}, 390--394\relax
\mciteBstWouldAddEndPuncttrue
\mciteSetBstMidEndSepPunct{\mcitedefaultmidpunct}
{\mcitedefaultendpunct}{\mcitedefaultseppunct}\relax
\EndOfBibitem
\bibitem[Davis and Chung(1982)Davis, and Chung]{davis1982mass}
Davis,~B.~F.; Chung,~K.~T. Mass-polarization effect and oscillator strengths for S, P, D states of helium. \emph{Physical Review A} \textbf{1982}, \emph{25}, 1328\relax
\mciteBstWouldAddEndPuncttrue
\mciteSetBstMidEndSepPunct{\mcitedefaultmidpunct}
{\mcitedefaultendpunct}{\mcitedefaultseppunct}\relax
\EndOfBibitem
\bibitem[Marinica \latin{et~al.}(2006)Marinica, Gaigeot, and Borgis]{marinica2006}
Marinica,~D.~C.; Gaigeot,~M.-P.; Borgis,~D. Generating approximate {Wigner} distributions using {Gaussian} phase packets propagation in imaginary time. \emph{Chemical physics letters} \textbf{2006}, \emph{423}, 390--394, Publisher: Elsevier\relax
\mciteBstWouldAddEndPuncttrue
\mciteSetBstMidEndSepPunct{\mcitedefaultmidpunct}
{\mcitedefaultendpunct}{\mcitedefaultseppunct}\relax
\EndOfBibitem
\end{mcitethebibliography}

\end{document}